\documentclass{statsoc}

\usepackage[a4paper]{geometry}
\usepackage{amsmath,graphicx}
\usepackage{natbib}

\RequirePackage[OT1]{fontenc}
\RequirePackage[colorlinks,citecolor=blue,urlcolor=blue]{hyperref}

\title[Cost-effectiveness analysis]{Statistical methods for cost-effectiveness analysis of left-truncated censored survival data with treatment delays}

\author[Khudyakov {\it et al.}]{Polyna Khudyakov}
\address{Harvard School of Public Health, Boston, MA, USA.}

\author{Li Xu}
\address{Harvard School of Public Health, Boston, MA, USA.}

\author{Ce Yang}
\address{Harvard School of Public Health, Boston, MA, USA.}

\author{Donna Spiegelman}
\address{Harvard School of Public Health, Boston, MA, and Yale School of Public Health, New Haven, CT, USA.}

\author{Molin Wang}
\coaddress{Molin Wang, Channing Division of Network Medicine, Harvard Medical School and Brigham and Women's Hospital, 180 Longwood Avenue and Department of Epidemiology and Biostatistics, Harvard T.H. Chan School of Public Health, Boston, Massachusetts 02115, USA \\
Email: stmow@channing.harvard.edu}
\address{Harvard School of Public Health, Boston, MA, USA.}

\begin{document}

\begin{abstract}
The incremental cost-effectiveness ratio (ICER) and incremental net benefit (INB) are widely used for cost-effectiveness analysis. 
We develop methods for estimation and inference for the ICER and INB which use the semiparametric stratified Cox proportional hazard model, allowing for adjustment for risk factors. 
Since in public health settings, patients often begin treatment after they become eligible, we account for delay times in treatment initiation. 
Excellent finite sample properties of the proposed estimator are demonstrated in an extensive simulation study under different delay scenarios. 
We apply the proposed method to evaluate the cost-effectiveness of switching treatments among AIDS patients in Tanzania.  
\end{abstract}


\section{Introduction}
Evaluating the cost-effectiveness of a clinical or public health intervention is a critical consideration for policy makers when making broad-based recommendations.  
In randomized clinical trials, especially in the pharmaceutical industry, it has become usual practice to include cost-effectiveness as a secondary outcome [\cite{Obrien}].
The most widely used measure of cost-effectiveness is the incremental cost-effectiveness ratio (ICER) [\cite{Laska, Laska2, Obrien2}], defined as the ratio of the expected difference in the costs between two alternative interventions over the lifespan and the expected differences in life expectancies. 
Because few studies can observe events over the entire lifespan, the restricted mean survival time (RMST), defined as the mean survival time within a restricted time period, say $[0,\eta]$, is the most widely used approach to calculate life expectancy. 
It is equal to the area under the survival curve up to time $\eta$ [\cite{Irwin,Zea}], and can be estimated by the Kaplan-Meier curve [\cite{ZT,ZT2}], Cox proportional hazards model [\cite{Dudley,Zucker}], the Aalen regression model with additive hazard function [\cite{Pagano}] or thorough parametric proportional hazard methods [\cite{Dudley}].

One drawback of the ICER is numerical instability when the denominator is close to zero, that is, when the intervention has very little effect. 
It is questionable whether cost-effectiveness is even of interest in such a situation. 
Nevertheless, to overcome this problem, the incremental net benefit (INB) was proposed, where the expected difference in medical costs is subtracted from the expected differences in life expectancies multiplied by the willingness-to-pay, i.e. to estimate how much the payer is ready to pay for one unit of health benefit, which we will denote as $\theta$ [\cite{INB,INB2}].

Our study was motivated by a study of the comparative effectiveness of switching from first line (ARV1) to second line (ARV2) antiretroviral drugs  in relation to all-cause mortality among patients attending HIV clinics in Dar es Salaam, Tanzania, between 2004 to 2012 [\cite{CH}]. 
In this study, most patients were not followed for the entire period of interest, 2004-2012, due to censoring. To estimate the survival functions, we use the Cox proportional hazard model because of its ability to deal with censored data without distributional assumptions and for its convenience in adjusting for covariates. 
In our motivating example, medical expenses were paid by the Tanzanian government, so the price depended only on the treatment type,  is the same for all patients within the same treatment group, and  did not change over time during the study period. 
In addition, treatment price did not influence patient choice, avoiding a situation where patients switch from one medication to another or stop using it at all because it is too expensive. 
In these circumstances, it is reasonable to assume that there was no informative censoring caused by treatment price, and thus standard survival analysis techniques are valid. 
\textcolor{black}{To allow a possible violation of the proportional hazard assumption, we use a Cox proportional hazard model stratified by treatment type.}

Cost effectiveness measures are absolute measures of effect, and, as such, are sensitive to the undelying risk factor distribution [\cite{Tyler2}]. 
In addition to presenting methodology for the estimation of cost-effectiveness measures adjusted for confounders and their distribution in the observed data, our methodology allows for the standardization of these measures to hypothetical or real covariate distributions of interest, such as to a country's general population or another standard of interest for public policy. 
To ensure external generalizability, this distribution must be carefully chosen [\cite{Tyler2}]. 
In addition, as exemplified in our motivating data example, treatment initiation delays following ARV2 eligibility must be considered. 
These delays occur when the prescribed treatment is unavailable, or too toxic, or when patients otherwise do not feel comfortable with immediately starting the treatment. 
In resource-limited settings, these are all common occurrences. 
For example, only 1.4\% of the eligible patients in our large cohort of patients undergoing AIDS  treatment initiated ARV2 at the time of becoming eligible. 
This phenomenon is also observed for other types of treatments, especially when the treatments are toxic. 
To our knowledge, accounting for treatment delays in the calculation of ICER is a novel development. 
In this paper, we develop methods for adjusting for delay time, and explore the impact of different delay scenarios on measures of cost-effectiveness, including the hypothetical one where no one experiences a delay.

This paper is organized in five sections. 
In Section 2, we set out notation for the problem and develop the methods. 
In Section 3, we describe an extensive simulation study. 
In Section 4, we apply the methodology to the data from HIV clinics in Dar es Salaam, Tanzania for estimation and inference about cost-effectiveness for ARV2 switching. 
Finally, in Section 5, we summarize and discuss the practical importance of our findings.

\section{Methods}
\label{s:methods}

\subsection{Measures of cost-effectiveness}

\textcolor{black}{We aim to determine the cost-effectiveness of Treatment $j$, $j=2, \ldots, J$, compared to the existing treatment, Treatment 1, over a follow-up period from 0 to $\eta$.
We denote survival times for the treatment group $j$ as $T_j$, and we assume that Treatment $j$ has a constant cost per unit of time, $c_j$, $j = 1, \ldots, J$. 
The RMST [\cite{Tian,Zea}] for treatment group $j$ is defined as $E(\min(T_{j}, \eta))$.  
Costs accumulated up to time $\eta$ will be estimated by the product of the RMST and cost of treatment per unit of time per person. 
If all patients start the treatment at time 0, the ICER is
$$
ICER_j=\frac{c_j E(\min(T_{j},\eta))-c_1 E(\min(T_{1},\eta))}{E(\min(T_{j},\eta))-E(\min(T_{1},\eta))}.
$$}
The INB for Treatment $j$ is
\[
{INB}_j(\theta)=\theta\{E(\min(T_{j},\eta))-E(\min(T_{1},\eta))\}-\{c_jE(\min(T_{j},\eta))-c_1E(\min(T_{1},\eta))\},
\]
where $\theta$ is the willingness-to-pay.

The key elements in both cost-effectiveness measures are the RMSTs, $E(\min(T_{j},\eta))$, $j=1, \ldots, J$. 
In what follows, we develop several options for their estimation and inference.

\subsection{Estimation and inference for the RMST}
\label{Cox}

In the study that motivated this research, survival times may not be observed due to censoring and the survival time for the $i$-th individual in the $j$-th treatment group depends on covariates, ${\boldsymbol{X}}_{ij}$. 
In addition, the treatment effect for the hazard function for death may not necessarily be multiplicative in the covariates. 
Therefore, we apply a stratified Cox proportional hazard model
$\lambda_{ij}(t)=\lambda_{0j}(t)e^{{\boldsymbol{\beta}}^T{\boldsymbol{X}}_{ij}}$, where $\boldsymbol\beta$ is a $p$-vector of unknown regression coefficients and $\lambda_{0j}(t)$ is baseline hazard function of unspecified form, to estimate $\boldsymbol{\beta}$ by the maximum partial likelihood method [\cite{KP}]. 
Then, we estimate the treatment group-specific cumulative baseline hazard functions $\Lambda_{0j}(t) = \int_0^t \lambda_{0j}(u)du$ using the Breslow type estimator [\cite{Breslow}]. 
Although here, for simplicity, we assume that the regression coefficients, $\boldsymbol{\beta}$, are the same for all treatment groups, the methodology also applies to the more general situation by adding interaction terms to the stratified Cox model above. 
We denote the observed follow-up time for individual $i$ in group $j$ by $\widetilde T_{ij} = \min(T_{ij},C_{ij})$, where $T_{ij}$ and $C_{ij}$ are the survival and censoring times of  individual $i$ in group $j$.
In the motivating example, the time scale is time since eligibility for 2nd line, within the maximum follow-up duration of $[0,\eta]$.

If all patients start treatment as soon as they become eligible, that is, all patients start the treatment at baseline, $t=0$, then the RMST for group $j$, $\mu_j=E(\min(T_{j},\eta))$, $j=1,\ldots,J$, can be estimated conditional on ${\boldsymbol{X}}_{ij}={\boldsymbol{x}}_{ij}$ as follows [\cite{Zucker}]
\[
\hat\mu_j ({\boldsymbol{x}}_{ij})=\int_0^\eta \widehat{S}_{j} (t|{\boldsymbol x}_{ij})dt,
\]
where $\widehat{S}_{j} (t|{\boldsymbol x}_{ij})=\exp\{-\int_0^t \widehat\lambda_{j}(t|{\boldsymbol x}_{ij})dt\}$ is the estimator of the survival function from the stratified Cox model $\lambda_{j}(t|{\boldsymbol x}_{ij}) = \lambda_{0j}(t) \exp({\boldsymbol{\beta}}^T {\boldsymbol{x}}_{ij})$, and $i=1,\ldots,n_j$ where $n_j$ is a number of participants on treatment $j$.
As shown in \cite{Zucker}, the variances of $\hat\mu_j (\boldsymbol{x})$ can be defined by methods similar to those used by \cite{AG}, for working out the asymptotic behavior of the Breslow estimator and the approximation
$\sqrt{n}({\boldsymbol{\hat{\beta}}}-{\boldsymbol{\beta}})\rightarrow \boldsymbol{N}(\boldsymbol{0},{\bf{\Sigma}}^{-1})$, where $\bf{\Sigma}$ is the limiting value of the partial likelihood information matrix normalized through division by $n=\sum_{j=1}^J n_j$.

As discussed above, in the motivating data example studying the effectiveness of switching from ARV1 to ARV2 and in other large-scale settings, patients do not always start their recommended treatment right after it is prescribed. 
We define the treatment delay time of the $i$th individual in group $j$ as $\delta_{ij}$.
The survival function for those in treatment group $j$ can only be estimated for $t>\delta_j$, where $\delta_j=\min\{\delta_{1j}, \ldots, \delta_{n_jj}\}$.
The group-specific baseline cumulative hazard functions can be estimated using the Breslow type estimator
\begin{equation}\label{LAMBDA}
\widehat{\Lambda}_{0j}(t)=
\sum_{\{p: T_{j(p)}\le t\}}\Big[\sum_{i=1}^{n_j}Y_{ij}(T_{j(p)})\exp({\boldsymbol{\hat{\beta}}}^T \boldsymbol{X}_{ij})\Big]^{-1},
\end{equation}
where $t>\delta_{j}$, $T_{j(p)}$ is the $p$th event time in group $j$, and $Y_{ij}(T_{j(p)})=I_{\{\widetilde T_{ij}\geq T_{j(p)}\}}$.
It follows that $\widehat{S}_j (t | \boldsymbol{X}_{ij}; T>\delta_j) = \exp \{ -\widehat\Lambda_{0j}(t) \exp({\boldsymbol{\hat\beta}}^T \boldsymbol{X}_{ij}) \}$.
The asymptotic behavior of the Breslow estimator of baseline hazard functions for the stratified case was derived by \cite{AG}.

As in our motivating example, all patients were on Treatment 1, here ARV1, at baseline, the time of ARV2 eligibility. 
In what follows, we assume that patients were in group 1 when they were found eligible for another treatment $j$. 
In the observed data, some patients switch to Treatment $j$ right after becoming eligible and others later.
We propose three ways to calculate RMST in the presence of these delay times: Scenario STRT (an abbreviation of ``start") estimates RMST under left truncation; 
that is, follow-up time starts after the baseline and the RMST in this scenario represents the mean survival time after the start of followup for participants who survive to a truncation time $r$. 
Scenario DLY (an abbreviation of ``delay") assumes one common delay time, and Scenario DST (an abbreviation of ``distribution") takes all observed (or hypothetical) delay times into account. 
Scenario DLY is applicable when there is interest in the cost-effectiveness under a given treatment delay time, which could provide an estimate of the cost-effectiveness under imperfect treatment protocol administration, or at perfect protocol administration such as zero delay time. 
Scenario DST applies to a study population with a range of delay time or to a hypothetical population's delay time distribution. 
For all three scenarios, we will evaluate how the definitions of RMSTs affect both of the cost-effectiveness measures considered in this paper, the ICER and INB.

\subsection{RMST for patients who survive to a fixed time $r$ (STRT) }
\label{S1}

\begin{figure}
\begin{center}
\includegraphics[height=2.2 in]{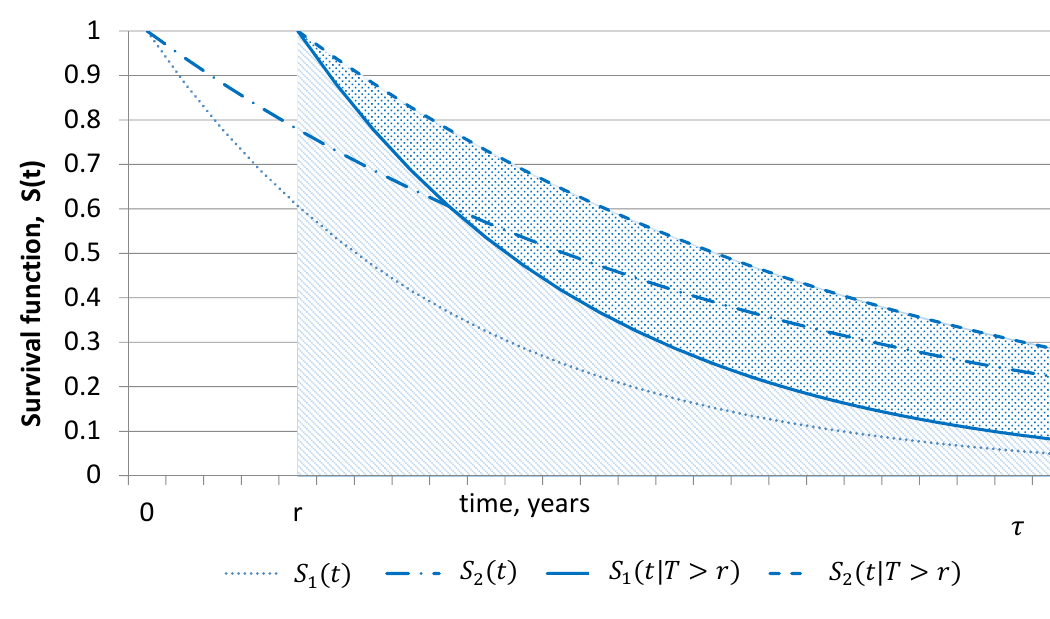}
\end{center}
\caption{\label{strt1} RMST for Treatments 1 and 2 under Scenario STRT (i.e., for patients who survive to a fixed time $r$). The shaded area under the solid curve corresponds to the conditional RMST for Treatment 1, $E(T_1-r|T_1>r)$, and the shaded area under the dashed curve corresponds to the conditional RMST for Treatment 2, $E(T_2-r|T_2>r)$.}
\end{figure}

In this scenario, we consider conditional RMSTs calculated for each treatment $j$ conditional upon survival from the baseline to the truncation time $r$, i.e., $E(T_j-r|T_j>r)$. 
The survival probabilities for a case with two treatment groups (j=1,2) are illustrated on Figure \ref{strt1}.


Here, one might set the \textcolor{black}{pre-specified truncation time} to the mean delay time or any delay time in the range of ($\delta$, $\eta$), where $\delta=\max\{\min\{\delta_{ij}\}_{i=1}^{n_j}\}_{j=1}^J$; i.e., $\delta$ is the maximum of the minimum delay times observed in each group. 
Also, note that $r$ has to be greater than $\delta$, because otherwise in some groups there would be no data to estimate the survival function from $r$ to $\delta$. Therefore, in what follows we assume $r>\delta$.

We can define the RMST under scenario STRT for individuals with $\boldsymbol{X} = \boldsymbol x$ as $\mu_{j}^{STRT}({\boldsymbol x},r) = E(T_{j}-r|T_{j}>r, \boldsymbol X=\boldsymbol x)$, which can be estimated by
$$
\hat \mu_{j}^{STRT}({\boldsymbol{x}},r)=\int_r^\eta
\widehat{S}_{j} (t|{\boldsymbol{X}=\boldsymbol{x}},T_{j}>r)dt=\frac{1}{\widehat{S}_{j} (r|{\boldsymbol{x}})}\int_r^\eta
\widehat{S}_{j} (t|{\boldsymbol{x}})dt.
$$
Alternatively, RMST could be estimated to correspond to the observed covariate distribution, ${\boldsymbol{\chi}}=\{\{{\boldsymbol{X}}_{ij}\}_{i=1}^{n_j}\} _{j=1}^{J}$, in the study population as
$\bar\mu_{j}^{STRT}({\boldsymbol{\chi}},r) = \frac1{n} \sum_{k=1}^{J} \sum_{i=1}^{n_k} \hat\mu_{j}^{STRT}({\bf{X}}_{ik},r)$, where $n = \sum_{j=1}^J n_j$.
We could instead use an external covariate distribution of a population to which it is desired to generalize the ICER. 
\textcolor{black}{This will lead to
$\bar\mu_{j}^{STRT}({\boldsymbol{\chi}},r) = \sum_{q\in Q}Pr({\boldsymbol{X}} = {\boldsymbol{X}}_q) \hat\mu_{j}^{STRT} ({\boldsymbol{X}}_{q}, r)$, where $Q$ is the set of external variable values and $Pr(\cdot)$ is the proportion of each unique set in the overall population}.
\textcolor{black}{For presentational simplicity, we will denote $\bar\mu_{j}^{STRT}({\boldsymbol{\chi}},r)$ as $\bar\mu_{j}^{STRT,2}({\boldsymbol{\chi}},r)$ in Section \ref{S26}.}

Next, we present the variance estimator of $\bar \mu_{j}^{STRT}({\boldsymbol{\chi}}, r)$, extending the developments in \cite{Zucker}.
Here, we allow for left truncation, such as that induced by delay times. 
Denote $S_{j} (t|T>r,{\boldsymbol{X}})$ as $S_{j}^{(r)} (t|{\boldsymbol{X}})$.
The estimator of $S_{j}^{(r)}(t|{\boldsymbol{X}})$ is 
$\widehat S_{j}^{(r)}(t|{\boldsymbol{X}}) = \exp\{- e^{{\hat{\boldsymbol{\beta}}}^T {\boldsymbol{X}}} (\widehat\Lambda_{0j}(t) - \widehat\Lambda_{0j}(r))\}$, where $\widehat{\Lambda}_{0j}(t)$ is given by \eqref{LAMBDA}.
Similar to \cite{Zucker}, we can show that $\bar\mu_{j}^{STRT}({\boldsymbol{\chi}}, r)$, with ${\boldsymbol{\chi}} = \{ \{{\boldsymbol{X}}_{ij}\}_{i=1}^{n_j}\} _{j=1}^{J}$, is an asymptotically normally distributed, consistent estimator with
mean $\mu_{j}^{STRT}({\boldsymbol{\chi}}, r) = n^{-1} \sum_{k=1}^{J} \sum_{i=1}^{n_k}\mu_j^{STRT}(X_{ik}, r)$, and estimated asymptotic variance  
\begin{equation}
\label{Var_mu}
\widehat{Var}(\bar{\mu}_{j}^{STRT}({\boldsymbol{\chi}},r))=n_j^{-1}{\Omega}_j^{(r)}({\boldsymbol{\chi}})+n^{-1}{\boldsymbol{{\Psi}}}_j^{(r)T}({\boldsymbol{\chi}})
{\boldsymbol{\widehat{\Sigma}}}^{-1}
{\boldsymbol{{\Psi}}}_j^{(r)}({\boldsymbol{\chi}}),
\end{equation}
where ${\Omega}_j^{(r)}({\boldsymbol{\chi}})$ and ${\boldsymbol{\Psi}}_j^{(r)}({\boldsymbol{\chi}})$ are defined in Appendix A, and ${\boldsymbol{\widehat{\Sigma}}}$ is defined in Section \ref{Cox}. 
See Web Supplementary Appendix 1 for technical details.

\subsection{RMST for a given fixed delay time $a$ (DLY)}
\label{S2}

\begin{figure}
\begin{center}
\includegraphics[height=2.2 in]{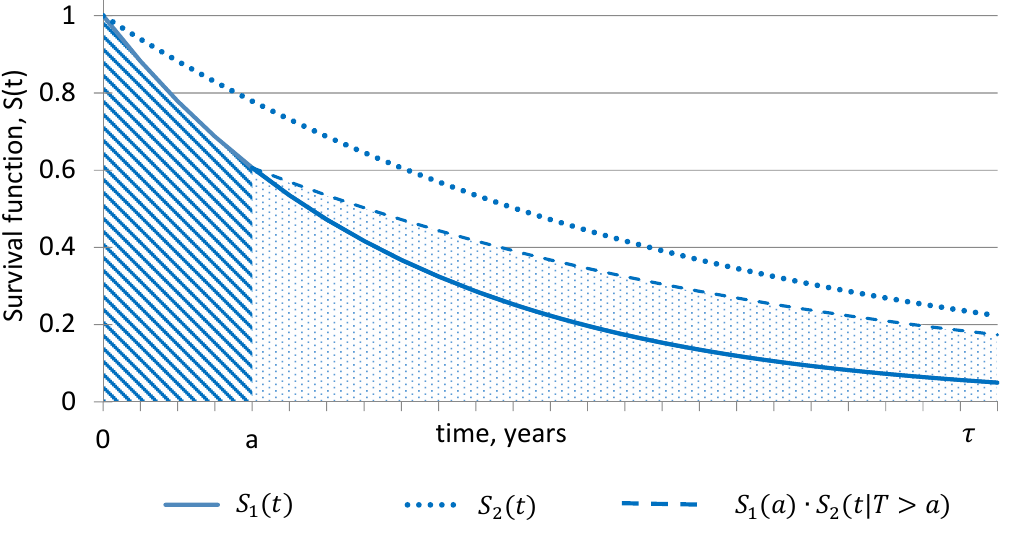}
\end{center}
\caption{\label{dly2} RMST for Treatments 1 and 2 under Scenario DLY (i.e., for a fixed delay time $a$). The area shaded by inclined lines is the part of RMST before the switch to Treatment 2, and the area under the dashed curve corresponds to the part of RMST after switching to Treatment 2. The area under solid line correspond to RMST under Treatment 1.}
\end{figure}

In scenario DLY, we aim to consider the full period of follow-up starting from time of eligibility for Treatment $j$ ($j=2,...,J$) to the end of follow-up, assuming a fixed time, $a$, from becoming eligible for Treatment $j$ to starting the treatment. 
Therefore, for patients who choose to switch from Treatment 1 to Treatment $j$, where $j$ is a treatment option, $j = 2, \ldots, J$, we estimate their RMST as the sum of two components: (1) RMST for Treatment 1 before the delay time $a$, and (2) RMST for Treatment $j$ for patients who survive to time $a$, at which point they are initiated to Treatment $j$.
Note that $a=0$ means no delay.  
The survival functions for Treatments 1 and 2 are illustrated on Figure \ref{dly2}.


For a fixed, pre-specified
value of the covariates, ${\boldsymbol{X}}={\boldsymbol{x}}$, we estimate RMST under scenario DLY as:
\[
\widehat{\mu}_{j}^{DLY}({\boldsymbol x},a)=\int_0^{a}\widehat{S}_1(t|{\boldsymbol{x}})dt+\frac{\widehat{S}_1(a|{\boldsymbol{x}})}
{\widehat{S}_j(a|{\boldsymbol{x}})}
\int_{a}^{\eta}\widehat{S}_j(t|{\boldsymbol{x}})dt.
\]
\textcolor{black}{We denote the second term as
\[
\widehat{\mu}_{j}^{DLY, 2}({\boldsymbol x},a) = \frac{\widehat{S}_1 (a|{\boldsymbol{x}})}
{\widehat{S}_j (a|{\boldsymbol{x}})}
\int_{a}^{\eta} \widehat{S}_j (t|{\boldsymbol{x}}) dt.
\]
As in scenario STRT, the RMSTs can also be estimated as \small{$\bar\mu_{j}^{DLY}({\boldsymbol{\chi}}, a) = \frac1{n} \sum_{k=1}^{J} \sum_{i=1}^{n_k} \hat\mu_{j}^{DLY} ({\boldsymbol{X}}_{ik}, a)$}, where $\boldsymbol{\chi}$ could be the covariate distribution observed in the study population, or in some other population standard.
The term corresponding to interval $[a, \eta]$ is \small{$\bar\mu_{j}^{DLY, 2}({\boldsymbol{\chi}}, a) = \frac1{n} \sum_{k=1}^{J} \sum_{i=1}^{n_k} \hat\mu_{j}^{DLY, 2} ({\boldsymbol{X}}_{ik}, a)$.}}

We define the asymptotic distribution for $\widehat\mu_{1}^{DLY}({\boldsymbol{X}},a)$ by extending \cite{Zucker}. 
The RMST for group 1 is given by
$\widehat\mu_{1}^{DLY}({\boldsymbol{X}},a)=\int_0^\eta \widehat S_1(t|{\boldsymbol{X}})dt$, with $S_1(t|{\boldsymbol{X}})=\exp\{-e^{{\boldsymbol{\beta}}^T{\boldsymbol{X}}}\Lambda_{01}(t)\}$.
The asymptotic distribution of RMST for subjects who switched to Treatment $j$, $\widehat{\mu}_{j}^{DLY}({\boldsymbol{X}},a)$, can be derived as the distribution of the sum of $\widehat\mu_{1}({\boldsymbol{X}},a)$ up to $a$, and of a second term defined as
$\widetilde S_{j}^{(a)}(t|X)=S_1(a|{\boldsymbol{X}})S_j(t|T>a,{\boldsymbol{X}}) = \exp \{-e^{{\boldsymbol{\beta}}^T{\boldsymbol{X}}}
\left[\Lambda_{01}(a)-\Lambda_{0j}(a) + \Lambda_{0j}(t)\right]\}$.
As derived in Web Supplementary Appendix 2, the variances for $\bar{\mu}_{1}^{DLY}({\boldsymbol{\chi}}, a)$ and $\bar{\mu}_{j}^{DLY}({\boldsymbol{\chi}}, a)$ can be estimated as in \eqref{Var_mu}, namely,
$\widehat{Var}(\bar{\mu}_{1}^{DLY}({\boldsymbol{\chi}}, a)) = n_1^{-1}{\Omega}_1({\boldsymbol{\chi}})+
n^{-1}{\boldsymbol{\Psi}}_1^{T}({\boldsymbol{\chi}}){\boldsymbol{\widehat{\Sigma}}}^{-1}
{\boldsymbol{{\Psi}}}_1({\boldsymbol{\chi}})$
and
$\widehat{Var}\left(\bar{\mu}_{j}^{DLY}({\boldsymbol{\chi}},a)\right)=n_1^{-1}{\widetilde\Omega}_1^{(a)}({\boldsymbol{\chi}})+
n_j^{-1}{\widetilde\Omega}_j^{(a)}({\boldsymbol{\chi}})+
n^{-1}({\boldsymbol{{\widetilde\Psi}}}_1^{(a)}({\boldsymbol{\chi}})+{\boldsymbol{{\widetilde\Psi}}_j^{(a)}}({\boldsymbol{\chi}}))^T
{\boldsymbol{\widehat{\Sigma}}}^{-1}({\boldsymbol{\widetilde\Psi}}_1^{(a)}({\boldsymbol{\chi}})+{\boldsymbol{\widetilde\Psi}}_j^{(a)}({\boldsymbol{\chi}}))$,
where ${\Omega}_1({\boldsymbol{\chi}})$, ${\boldsymbol{\Psi}}_1({\boldsymbol{\chi}})$, ${\widetilde\Omega}^{(a)}_j({\boldsymbol{\chi}})$ and ${\boldsymbol{\widetilde\Psi}}^{(a)}_j({\boldsymbol{\chi}})$ are defined in Appendix B.

\subsection{RMST for a given distribution of delay times (DST)}
\label{S3}

So far, although we have allowed for the RMST to be evaluated over the observed distribution of covariates in the study or any other covariate distribution, $\bar\mu_j^{STRT}({\boldsymbol{\chi}},a)$ and $\bar\mu_j^{DLY}({\boldsymbol{\chi}}, a)$ are both evaluated at a pre-specified single delay time. 
In this sub-section, we further generalize the RMST to allow for its evaluation over the observed distribution of delay times or some other hypothetical distribution of delay times. 
We write the RMST as an average of the RMSTs over the delay time distribution, $\{\delta_{ij}\}_{i=1}^{n_j}$, where $n_j$ is the number of subjects in group $j$. 
In this case, the RMST in group $j$ for a population with covariate distribution $\boldsymbol{\chi}$ and delay distribution ${\boldsymbol{\Delta}} = \{ \delta_{lj}\}_{l=1}^{n_j}$ can be written as
\[
\bar\mu_j^{DST}({\boldsymbol{\chi}},{\boldsymbol{\Delta}})=\frac{1}{n_j} \sum_{l=1}^{n_j} \bar{\mu}_{j}^{DLY}({\boldsymbol{\chi}},\delta_{lj}),
\]
where
\[
\bar{\mu}_{j}^{DLY}({\boldsymbol{\chi}},\delta_{lj})=\frac 1{n}\sum_{k=1}^{J}\sum_{i=1}^{n_k}\left( \int_0^{\delta_{lj}}\widehat S_1(t|{\boldsymbol{X}}_{ik})dt+
\frac{\widehat S_1(\delta_{lj}|{\boldsymbol{X}}_{ik})}{\widehat S_j(\delta_{lj}|{\boldsymbol{X}}_{ik})}\int_{\delta_{lj}}^{\eta}\widehat S_j(t|{\boldsymbol{X}}_{ik})dt\right).
\]
\textcolor{black}{We denote the term corresponding to the interval after the delay time as
\[
\bar\mu_j^{DST, 2}({\boldsymbol{\chi}},{\boldsymbol{\Delta}}) = \frac 1{n_j} \sum_{l=1}^{n_j} \left( \frac{1}{n}\sum_{k=1}^{J}\sum_{i=1}^{n_k} \left( 
\frac{\widehat S_1 (\delta_{lj}|{\boldsymbol{X}}_{ik})}{\widehat S_j (\delta_{lj}|{\boldsymbol{X}}_{ik})} \int_{\delta_{lj}}^{\eta} \widehat S_j(t|{\boldsymbol{X}}_{ik})dt \right) \right).
\]}
The variance of $\bar{\mu}^{DST}({\boldsymbol{\chi}},{\boldsymbol{\Delta}})$ is similar to the variance in scenario DLY, but with the
addition of the covariance terms between $\bar{\mu}_{j}^{DLY}({\boldsymbol{\chi}},\delta_{lj})$ and $\bar{\mu}_{j}^{DLY}({\boldsymbol{\chi}},\delta_{hj})$, 
\footnotesize{
\begin{equation}
\label{Psi}
\begin{split}
Cov(\bar\mu_{j}^{DLY}({\boldsymbol{\chi}},\delta_{lj}),\bar\mu_{j}^{DLY}({\boldsymbol{\chi}},\delta_{hj}))=&
	n_j^{-1}\big[\widetilde\Omega_1^{(\delta_{lj},\delta_{hj})}({\boldsymbol{\chi}})+\widetilde\Omega_j^{(\delta_{lj},\delta_{hj})}({\boldsymbol{\chi}})\\
	&+
	({\boldsymbol{\widetilde{\Psi}}}_1^{(\delta_{lj})}({\boldsymbol{\chi}})+{\boldsymbol{\widetilde{\Psi}}}_j^{(\delta_{lj})}({\boldsymbol{\chi}}))^T
	{\boldsymbol{\widehat{\Sigma}}}^{-1}
	({\boldsymbol{\widetilde{\Psi}}}_1^{(\delta_{hj})}({\boldsymbol{\chi}})+{\boldsymbol{\widetilde{\Psi}}}_j^{(\delta_{hj})}({\boldsymbol{\chi}}) )\big],
\end{split}
\end{equation}
}
where $l,h=1,...,n_j$, $h\ne l$,  and $\widetilde\Omega^{(\delta_{lj},\delta_{hj})}({\boldsymbol{\chi}})$ and $\widetilde{\Psi}_j^{(\delta_{lj})}({\boldsymbol{\chi}})$ are defined in Appendix B.

We could also be interested in evaluating the cost-effectiveness for a hypothetical delay time distribution. 
For example, we could be interested in a scenario when half of patients were initiated upon eligibility, and the reminder followed an exponential distribution, $f(\delta)=\alpha e^{-\alpha\delta}$. 
The estimate of $\bar{\mu}^{DST}( {\boldsymbol{\chi}},{\boldsymbol{\Delta}} )$ in this case will be
$
\bar\mu_j^{DST}({\boldsymbol{\chi}},{\boldsymbol{\Delta}})={0.5}\bar{\mu}_{j}^{DLY}({\boldsymbol{\chi}},0)
+0.5\int_0^\infty \bar{\mu}_{j}^{DLY}({\boldsymbol{\chi}},\delta)\alpha e^{-\alpha\delta}d\delta.
$

\subsection{Estimation and inference for the ICER and the INB}
\label{S26}

\textcolor{black}{The ICER is estimated by
\[
\widehat{ICER}_j^{(\cdot)}({\boldsymbol{\chi}},\boldsymbol{\Delta})=\frac{c_j \bar\mu_j^{(\cdot,2)}({\boldsymbol{\chi}},\boldsymbol{\Delta})
	-c_1 \bar\mu_1^{(\cdot,2)}({\boldsymbol{\chi}},\boldsymbol{\Delta})}
{\bar\mu_j^{(\cdot,2)}({\boldsymbol{\chi}},\boldsymbol{\Delta})-\bar\mu_1^{(\cdot,2)}({\boldsymbol{\chi}},\boldsymbol{\Delta})} =
f(\bar\mu_1^{(\cdot,2)}({\boldsymbol{\chi}},\boldsymbol{\Delta}),\bar\mu_j^{(\cdot,2)}({\boldsymbol{\chi}},\boldsymbol{\Delta})),
\]
where $(\cdot)$ corresponds to one of the proposed scenarios: STRT, DLY or DST, and $\bar\mu_j^{(\cdot,2)}({\boldsymbol{\chi}},\boldsymbol{\Delta})$ could be replaced by $\bar\mu_j^{STRT,2}({\boldsymbol{\chi}},\boldsymbol{\Delta})$, $\bar\mu_j^{DLY,2}({\boldsymbol{\chi}},\boldsymbol{\Delta})$ or $\bar\mu_j^{DST,2}({\boldsymbol{\chi}},\boldsymbol{\Delta})$, calculated over a fixed, observed or hypothetical covariate distribution $\boldsymbol{\chi}$, and $\boldsymbol{\Delta}$ is a fixed scalar value for scenarios STRT and DLY and an observed or hypothetical delay time distribution for scenario DST.
See the technical details in Web Supplementary Appendix 3.}

We derived $Var(\widehat{ICER}_j^{(\cdot)}({\boldsymbol{\chi}},\boldsymbol{\Delta}))$ using the delta method, where
\newline
$
\sqrt{n}\left(\widehat{ICER}_j^{(\cdot)}({\boldsymbol{\chi}},\boldsymbol{\Delta})-ICER_j\right)\longrightarrow {N}\left(\boldsymbol{0},\nabla f(\bar{\mu}_j^{(\cdot,2)}({\boldsymbol{\chi}},\boldsymbol{\Delta}),\bar{\mu}_1^{(\cdot,2)}({\boldsymbol{\chi}},\boldsymbol{\Delta}))^T\Sigma^*\nabla f(\bar\mu_j^{(\cdot,2)}({\boldsymbol{\chi}},\boldsymbol{\Delta}),\bar\mu_1^{(\cdot,2)}({\boldsymbol{\chi}},\boldsymbol{\Delta}))\right),
$
$\Sigma^*$ is a $2\times 2$ matrix with the diagonal elements $\Sigma^*_{11}=Var(\bar\mu_1^{(\cdot,2)}({\boldsymbol{\chi}},\boldsymbol{\Delta}))$ and $\Sigma^*_{jj}=Var(\bar\mu_j^{(\cdot,2)}({\boldsymbol{\chi}},\boldsymbol{\Delta}))$, $j=2,...,J$, and off-diagonal elements $\Sigma^*_{1j}=\Sigma_{j1}^*=Cov(\bar\mu_1^{(\cdot,2)}({\boldsymbol{\chi}},\boldsymbol{\Delta}),\bar\mu_j^{(\cdot,2)}({\boldsymbol{\chi}},\boldsymbol{\Delta}))$
are in Appendices A and B.

The INB is estimated by
$$
\widehat{INB}_j^{(\cdot)}({\boldsymbol{\chi}},\boldsymbol{\Delta},\theta)=\theta(\bar\mu_{j}^{(\cdot,2)}({\boldsymbol{\chi}},\boldsymbol{\Delta})-
\bar\mu_{1}^{(\cdot,2)}({\boldsymbol{\chi}},\boldsymbol{\Delta}))-
(c_j\bar\mu_{j}^{(\cdot,2)}({\boldsymbol{\chi}},\boldsymbol{\Delta})-c_1\bar\mu_{1}^{(\cdot,2)}({\boldsymbol{\chi}},\boldsymbol{\Delta})),
$$
where $\bar\mu_j^{(\cdot,2)}({\boldsymbol{\chi}},\boldsymbol{\Delta})$ is replaced by $\bar\mu_j^{STRT,2}({\boldsymbol{\chi}},\boldsymbol{\Delta})$, $\bar\mu_j^{DLY,2}({\boldsymbol{\chi}},\boldsymbol{\Delta})$ or $\bar\mu_j^{DST,2}({\boldsymbol{\chi}},\boldsymbol{\Delta})$ calculated over a fixed, observed or hypothetical covariate distribution ${\boldsymbol{\chi}}$. 
The variance of the INB$_j$ is \\
$\widehat{Var}(\widehat{INB}^{(\cdot)}({\boldsymbol{\chi}},\boldsymbol{\Delta},\theta))=\widetilde c_1^2 \widehat{Var}(\bar{\mu}_1^{(\cdot,2)}({\boldsymbol{\chi}},\boldsymbol{\Delta}))+\widetilde c_j^2
\widehat{Var}(\bar{\mu}_j^{(\cdot,2)}({\boldsymbol{\chi}},\boldsymbol{\Delta}))-2\widetilde c_1\widetilde c_j\widehat{Cov}(\bar{\mu}_1^{(\cdot,2)}({\boldsymbol{\chi}},\boldsymbol{\Delta}),
\bar{\mu}_j^{(\cdot,2)}({\boldsymbol{\chi}},\boldsymbol{\Delta}))$, where $\widetilde c_j=\theta-c_j$ and $j=1, \ldots, J$.

The RMSTs, their estimates and the variances of the estimates under the three scenarios considered differ. 
It is of interest to understand how the choice of scenario under which the RMST is evaluated influences the estimates of the ICER and the INB, and inference about them. 
In Web Supplementary Appendix 3, we prove that if the RMST is calculated under scenarios STRT or DLY for a pre-specified covariate value ${\boldsymbol X}={\boldsymbol x}$ and the same value of $a$, then $ICER^{STRT}_j({\boldsymbol x},a)=ICER_j^{DLY}({\boldsymbol x},a)$.
This is because the RMST up to the time period before initiating the treatment,  which is  included in scenario DLY but not in scenario STRT, is the same between the treatment groups, and thus, is canceled out in both the denominator and numerator of   $ICER^{DLY}_j$. 
In addition, $INB^{DLY}_j({\boldsymbol x},a,\theta)=S_1(a|{\boldsymbol x})INB_j^{STRT}({\boldsymbol x},a,\theta)$. 
Therefore, when estimating the ICER or INB for a pre-specified value of the covariates and a fixed delay time, the STRT is recommended because of its computational ease. 

Estimation and inference of the RMST has been well-studied in the literature.
\cite{Karrison1987} first investigated the asymptotic properties of RMST with adjustment for covariates under piece-wise exponential models.
The covariate-adjusted RMST procedure was later extended by \cite{Zucker} to general stratified Cox models.
As discussed in \cite{Zucker}, the baseline cumulative hazard functions were estimated using the Breslow estimator [\cite{Breslow}].
Under standard regularity conditions and correct specification of the Cox model, the MLE of the regression coefficient is consistent, and the difference between the Breslow estimator for the baseline cumulative hazard and true baseline cumulative hazard function weakly converges to a zero mean Gaussian Process [\cite{Tsiatis1981}].  
Extending the asymptotic results for the Cox model in \cite{Anderson2012} to the stratified case, \cite{Zucker} proved consistency and asymptotic normality for estimators of the regression coefficients used for RMSTs.
Thus, as continuous functions of the RMST, the ICER and INB estimators are also consistent.

Finally, we comment that if the assumption of proportional hazards is justified for both the intervention and covariate effects, as is often the case [\cite{Tyler}], it will be more efficient to fit a single Cox  model that includes group indicators, ${G}_{j} =(G_{2}, \ldots, G_{J})$, where $G_{j}=1$ for treatment $j$ and $0$ otherwise, 
$j=2, \ldots, J$, as covariates in addition to the other covariates, ${\boldsymbol X}$. 
Then, the baseline survival function will be the same between treatment groups, and
$\widehat S_j(t|{\boldsymbol X})=\widehat S(t|\boldsymbol{G}_{j},\,{\boldsymbol X})$ for $j=1, \ldots, J$. 
All other calculations will be the same, including the definitions of the different scenarios for the RMSTs calculations.

\section{Simulation study}
\label{s:sim}

To evaluate the finite sample performance of the methods developed in Section 2, we conducted an extensive simulation study. 
In each simulated scenario, the number of simulation replicates was 1,000, with study sample sizes 1,000 and 10,000, and an
assumed  study duration of 10 years, in which we estimate the cost-effectiveness of switching from Treatment 1 to Treatment 2. 
The survival times were generated assuming constant baseline hazard functions, $\lambda_{01}=1$ per year and $\lambda_{02}=0.2$, 0.5 and 0.8 per year, corresponding to the following choices for the hazard ratio (HR) for Treatment 2 compared to Treatment 1: 0.2 (substantial benefit), 0.5 (as was found in \cite{CH}) and 0.8 (widely observed in epidemiologic studies). 
The censoring times were generated
from the exponential distribution with rate 0.01 per year. 
For simplicity, we considered one binary covariate $X\sim Ber(0.9)$ generated independently of treatment group, and set $\beta=-2$ for the effect of this covariate. 
The delay times for Treatment 2 were first generated from a standard uniform distribution, $d_{i2}\sim Unif(0,1)$, following the distribution observed in our data, where all eligible patients switched to Treatment 2 within a year after eligibility.
To investigate the influence of the proportion of people who experienced a delay, we considered scenarios with different percents of individuals
with switching delays, namely, 0\%, 10\% and 50\%. 
For costs, we used values provided by a recent World Health Organization (WHO) Technical report [\cite{HIV}], that gave the cost of ARV1 as \$115 per person per year and \$330 per person per year for ARV2. 
For the INB, we chose a value for $\theta$ that corresponded to the Tanzanian per capita GDP based on purchasing-power-parity (PPP); that was \$1,352 in 2008 (International Monetary Fund data, available at \url{http://www.imf.org/external/ns/cs.aspx?id=28}), which is the midpoint of the study period (2004-2012) in the example motivating this research.


The results for $\widehat{\text{RMST}}$ for each treatment calculated under scenarios STRT and DLY, their SEs and percent coverage of $95\%$ CIs are presented in Table 1, while Table 2 summarizes the results for the $\widehat{ICER}$ and $\widehat{INB}$, together with the empirical coverage probabilities of the asymptotic 95\% confidence intervals (CIs). 
The results for scenario DST are presented in Table 3.

In the simulation study, we considered RMST estimates under the three scenarios STRT, DLY and DST. 
Note that with sample size 10,000 and 50\% delays, there were 5,000 observed delay times. 
Under the DST scenario, we need to calculate the RMST over all delay times. To reduce the heavy computational burden in this case, we estimated RMSTs at  10 distinct delay time values, i.e. $d_{i2}$ took values from 0.05 to 0.95, allowing for three options for the distribution of the delay times: uniform, right-skewed  and left-skewed as presented in Web Supplementary Figure 1.
Since the data for the simulations were generated from the Cox model with a constant baseline hazard function, the survival times have an exponential distribution, and therefore, in each of the simulation scenarios considered, we were able to derive the theoretical value of the RMSTs, ICERs and INBs implied by the simulation design (see Web Supplementary Appendix 4 for details), and compare them to the estimated values to assess finite sample bias. 
As presented in Web Supplementary Figure 2, as $\eta$ increases the values of ICER under scenario STRT go to their limiting values of \$383.75, \$545 and \$1,190 for HR=0.2, 0.5 and 0.8, respectively. In this simulation study, the limiting values for scenario STRT and DLY coincide, but this is not necessarily the case for the data settings other than in this simulation study.

As shown in Tables \ref{tab1}-\ref{tab3}, even for a sample size 1,000, the RMST estimates had little bias. 
\textcolor{black}{However, the estimation of the ICER and its variance can be problematic in a small sample when the treatment effect is small; specifically, the ICER estimation did not perform well when HR $= 0.8$ or closer to 1 and the sample size was 1000.
This problem is likely caused by the fact that ICER is a ratio in which, when the treatment effect is small, the denominator could be close to zero.}
In the simulation study, the instability of  $\widehat{\text{ICER}}$ was eliminated by increasing the sample size to 10,000. 
$\widehat{\text{INB}}$ performed well in all scenarios considered. 
The estimators exhibited the least finite sample bias in the absence of any delay, likely because left truncation does not occur. 
The greater the percentage of delayed initiation, the more bias there was. Empirical coverage probabilities were close to the desired 95\%, even for a sample size of 1,000, indicating that the proposed variance estimators are of high quality. 
Similar good results were found when we set the sample size as $n=500$.
See Tables 1 and 2 in Appendix 5 of the supplementary materials for details.
Moreover, we additionally considered simulation settings with higher and lower baseline hazards, leading to 0 - 76\% censoring rates and 0 - 62\% proportions missing Treatment 2 due to death, among those delayed in Treatment 2 initiation.
See Tables 3 - 5 in Appendix 5 of the supplementary materials for results as well as Table 6 in Appendix 5 of the supplementary materials for a summary of censoring rates and proportion of patients not receiving Treatment 2 due to delay across all the simulation settings.
The relative biases of RMST estimates ranged from 0.0\% to 2.3\%, and the relative biases of ICER and INB estimates ranged from 0.0\% to 10.3\% across these simulation settings, demonstrating good finite sample performance in a wide range of scenarios which may be encountered in practice.

\section{Example: Cost-effectiveness of switching to ARV2 in a large HIV/AIDS treatment program in Dar es Salaam, Tanzania}

The data motivating this research were collected by the Harvard-Management and Development for Health collaboration with the United State Government, who together have implemented the President's Emergency Plan for AIDS Relief (PEPFAR) program in Dar er Salaam, Tanzania. Research data are not shared. 
Data were available on clinical, laboratory and pharmacy characteristics of nearly 7,000 pediatric and nearly 110,000 adult patients between 2004, when the Government of Tanzania began providing ART to HIV-infected patients, until 2012. 
High rates of antiretroviral failure (from 11\% to 24\%) have been observed with much concern, and resistance to ARV1 has been noted, with ARV1 treatment failure reported in approximately 70\%-90\% of patients [\cite{C8, C9, C10}]. 
Although this is the recommended practice, the reported rates of switching to ARV2 in the presence of ARV1 failure have been reported to be low (2.4\%-30\%) [\cite{C1, C2, C3}]. 
However, a number of studies have demonstrated that switching to the 2nd line reduces mortality [\cite{C13,C14}.] 
Here, we are interested in estimating the cost-effectiveness of the switch to ARV2 in this large-scale, resource-limited setting.
Web Supplementary Figure 3 describes the data flow among ARV2-eligible patients in the program. 
Among the 1,455 ARV2-eligible patients, only 687 switched, and only 20 (1.4\%) of them did so immediately after eligibility has been ascertained. 
We applied the methodology described in Section 2 to estimate the RMSTs, ICERs and INBs under scenarios STRT, DLY and DST. 
The outcome in our study is all-cause  mortality up to 6 years following ARV2 eligibility. To examine the impact of the study duration on the cost-effectiveness measures, we conducted the analysis over  5, 5.5 and 6 years duration. 
The mean value of the delay times in the group of patients who switched to the 2nd line ART was 0.9 years, so we set $r=0$ and $0.9$ years in scenario STRT and $a=0$ and $0.9$ in scenario DLY. 
The actual distribution of delay times in the group of participants who switched to the 2nd line is presented in Web Supplementary Figure 4.


Following \cite{CH}, we controlled for the following potential confounders, measured at the time of being eligible for ARV2: gender, tuberculosis (TB) history (yes, no), HIV stage at second line eligibility, marital status, age at eligibility ($<$30, 30-$<$40, 40-$<$50, 50+ years), BMI ($<$18.5, 18.5-$<$25, 25-$<$30, 30+ kg/$m^2$), hemoglobin ($<$7.5, 7.5-$<$10, 10+ g/dl), CD4 ($<$50, 50-$<$100, 100-$<$200, 200+ cells/$mm^3$), use of cotrimoxazole (yes, no), facility level (hospital, health center, dispensary), district of Dar es Salaam, and visit adherence (proportion of days late for scheduled clinic visit). 
In addition, we controlled for the delay times (time since ARV2 eligibility to the ARV2 initiation) among those  who switched to ARV2; for those remaining on ARV1, this value was set to 0. 
We have also included a binary time-varying variable indicating before and after switching to ARV2 as a stratification factor in the Cox model. 
In Web Supplementary Table 1, we present the basic characteristics of the study population at the time of ARV2 eligibility. 
To show the influence of the covariate distribution in Web Supplementary Table 1, we present the results of analysis in which we estimated measures of cost-effectiveness for different covariates distributions: the observed covariate distribution, the median values of covariates and the covariate values of sicker participants, defined as females with BMI less than 18.5 who were 50+ years old at eligibility. 
For this group of sicker participants, all other covariates took values as observed in the population.

The hazard ratio (95\% CI) for switching to ARV2 versus remaining on ARV1 following ARV2 eligibility, adjusted for the potential confounders described above was 0.57 (0.33, 0.98). 
The multivariate-adjusted estimates of the RMSTs and the ICER, based on averaging over the observed distribution of all the potential confounders mentioned above, with no delay, a fixed delay time of 0.9 years, and the observed distribution of the delay times, are presented in Table \ref{rdata}. 
The values of the RMSTs were different for each scenario considered except for the RMST in group 1 under scenarios DLY and DST, when the formulas for both scenarios coincide due to the absence of delays. 
The estimated RMSTs under scenario STRT are smaller than other RMSTs values, because they were conditional on survival to 0.9 years and not including the survival duration from baseline to 0.9 years.


The ICER estimates and their $95\%$ CIs were relatively close between scenarios STRT and DLY for each follow-up period when evaluated at the observed covariate distribution. 
However, the ICERs in these two scenarios were not identical, as shown theoretically in Web Supplementary Appendix 3, because we were averaging over covariates distribution instead of evaluating at a fixed covariate value as in Web Supplementary Appendix 3 where they are identical. 
In Scenario DST, the estimated ICERs were bigger than in scenarios STRT and DLY due to the smaller difference between RMSTs in this scenario, and the $95\%$ CIs were wider than in scenarios STRT and DLY due to the additional variation introduced by taking into account the full delay time distribution. 
Benefit was most clearly shown for patients who were sicker at the time of eligibility. 
Importantly from a practical point of view, for the patients who were sicker, delays at the mean of 0.9 years led to poorer ICERs, 23\% greater than if the recommended policy of no delay would have been implemented.

\begin{figure}
\begin{center}
\includegraphics[height=3.1 in]{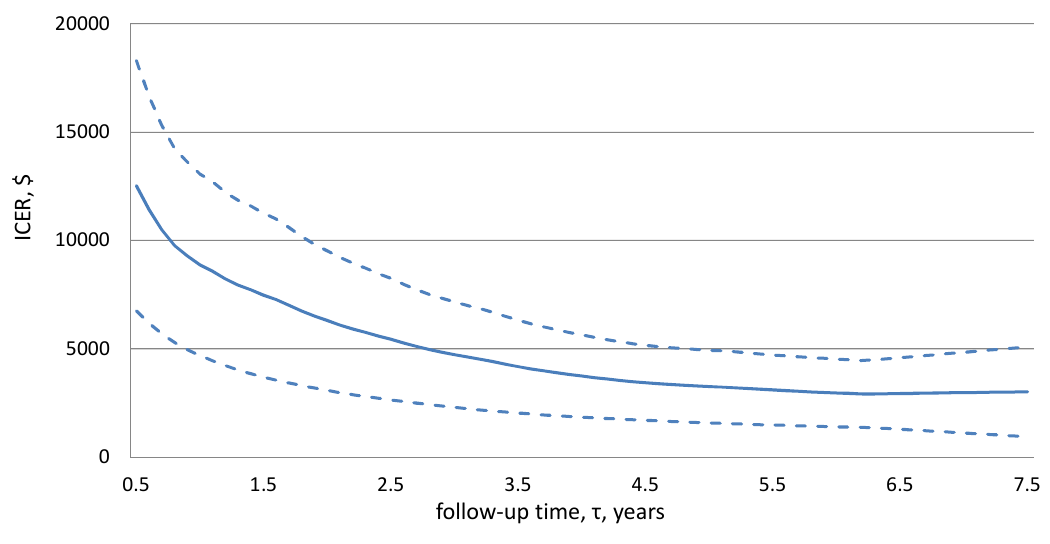}
\end{center}
\caption{\label{ICER_tau} $\widehat{\text{ICER}}$ (95\% CI) over a range of follow-up periods, $\eta$, 
under Scenario STRT (i.e., for patients who survive to a fixed time $r$; assuming the special scenario $r=0$), adjusted to the observed covariate distribution, Dar es Salaam, Tanzania, 2004-2012.}
\end{figure}

To illustrate the dependance of ICERs on the follow-up period duration, we calculated the ICERs and their $95\%$ CI for different values of follow-up time, $\eta$, under scenario STRT assuming immediate switching to ARV2 for those who became eligible, i.e. $r=0$. 
The results presented in Figure \ref{ICER_tau} show that cost-effectiveness decreases with the increasing duration of follow-up. 
There is a slight increase in the ICER when the follow-up period was greater than 6.5 years, which is likely due to numerical instability caused by sparse data towards the end of follow-up.

\begin{figure}
\begin{center}
\includegraphics[height=3.2 in]{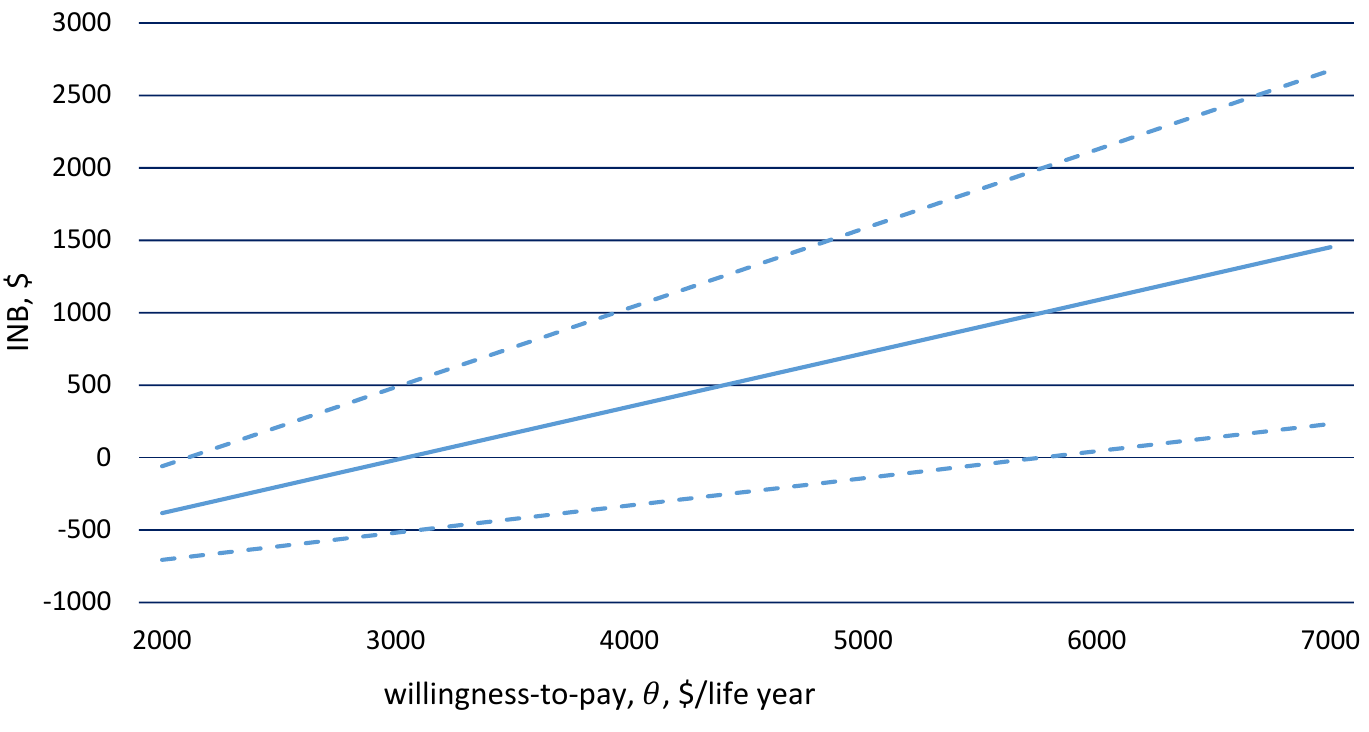}
\end{center}
\caption{ $\widehat{\text{INB}}$ (95\% CI) over a range of willingness-to-pay values, $\theta$, 
under Scenario STRT (i.e., for patients who survive to $0.9$ years), Dar es Salaam, Tanzanie, 2004-2012. Estimated RMST for ARV1 is $\bar{\mu}_1^{STRT}$=4.64 years and estimated RMST for ARV2 is $\bar{\mu}_2^{STRT}$=5 years, adjusted to the observed covariate distribution.}\label{INBf}
\end{figure}

As in the simulation study, the INB was calculated assuming that willingness-to-pay, $\theta$, was equal to the Tanzanian per capita GDP, \$1,352, in 2008 (the mid point of the follow-up period 2004-2012).
The relationship between the INBs under the different scenarios considered were similar to those of the ICER. 
The INBs for scenarios STRT and DLY were similar, and the INB for Scenario DST was much worse. 
As it was also demonstrated by the ICERs, sicker patients have a positive INB, indicating that switching to ARV2 was beneficial for sicker patients, but this benefit was estimated with a great deal of uncertainty, with the lower 95\% bound covering the zero value. 
We also present a plot of the multivariate-adjusted INB and its 95\% confidence interval in Scenario STRT for 6 year follow-up period where the willingness-to-pay takes  values from \$0 to \$13,000 (Figure \ref{INBf}). 
We can see that the point estimate of INB becomes positive when willingness-to-pay, $\theta$, was \$3,000 or more; 
however, the variation was high and the difference in willingness-to-pay needed to be more than \$6,000 for a year of life to provide a CI that does not include negative values.

This analysis showed that switching to ARV2 increased the RMST, since the difference between RMSTs in the two treatment groups was always positive under all scenarios considered. 
Because ARV2 is three times more expensive that ARV1, in all cases and scenarios considered except when a sicker population was defined, one additional year of life will cost 1.5 times more than the Tanzanian per capita GDP (\$1,352). 
Hence, ARV2 switching appeared to be, in general, not cost effective in this large-scale public health setting, even when the guidelines are followed perfectly, and all patients are initiated as soon as they are eligible.

\section{Discussion}
\label{s:discuss}

In this paper, we provided a solution to the problem of  cost-effectiveness analysis between two or more treatments when some subjects start the treatment of interest at a time after which the treatment is prescribed or at some time after they become eligible. 
In addition to permitting convenient and stable methods for adjusting for a large number of covariates, our methodology provides several options for accounting for an observed or hypothetical delay time distribution. 
Since, in reality, many patients do not start  treatments right after they are prescribed by a doctor, our methods may provide a more relevant estimate of cost-effectiveness than those that cannot account for the period before the patient starts to take the prescribed drug, and instead assume all patients start at the time of eligibility. 
\textcolor{black}{We presented three scenarios in order of complexity, STRT, DLY, and DST, and built technical details towards the DST scenario, the most general consideration of the three.
In the STRT scenario, only patients who survive up to the time $r$ at which point Treatment $j$ is initiated are followed for their subsequent survival times.
While patients may have their own delay times, when calculating the RMST, we let $r > \delta$, the maximum of the minimum delay times observed in each treatment group. 
Next, building upon the STRT scenario, in the DLY scenario, we described a situation where patients receiving the first line treatment may experience a delay time after becoming eligible for the second line treatment before being able to switch to the second line treatment.
For simplicity, a fixed delay time is assumed for everyone there.
Note that the entire time period from $t=0$ is considered in the DLY scenario while only the time period after $r$ is considered in the STRT scenario.
The DLY scenario later helped develop, and could be viewed as a special case of, the DST scenario, where delay times were allowed to be different among participants.}
We illustrated our methodology in a motivating example of substantial public health importance. 
While in our motivating example, the delay times were observed only in one group, the proposed approach can be also applied to a situation where patients experience delays in every treatment group. 
The computing software was coded using the SAS programming language, where the ``phreg" procedure was used to obtain the Breslow-type estimator in the Cox model.

The methodology presented allows for an assessment of cost-effectiveness assuming a fixed delay time, including, importantly, the optimal value of 0, as well as under an observed or hypothetical distribution of delay times. 
Thus, cost-effectiveness can be assessed as it would be expected to be observed in other program settings, or under hypothesized scenarios, instead of only according to the delay time distribution observed in the patient population providing the effectiveness estimates.

In the motivating study, the choice of the assumed delay time scenario had a substantial impact on the results. 
\textcolor{black}{The choice of the scenarios, covariate distribution and delay time distributions under which the ICER and INB are calculated are very important components of the analysis and should be determined by research and program needs.
If interest is in cost-effectiveness among those who survived to a point in time after becoming eligible for the second line treatment, scenario STRT should be used, and for a situation where the decision maker wants to choose the maximum allowable delay time for switching to the new treatment with cost-effectiveness, scenario DLY should be used. 
With scenario DST, a researcher can obtain externally generalizable estimates relevant to those observed in the study, or those that are expected to be observed in some broader group.}
\textcolor{black}{To some extent, our methodology may be helpful to policy makers who want to identify the best time for patients to start treatment by considering different delay times or delay time distributions to satisfy a pre-specified ICER value.
Moreover, our work elucidates further directions for the development of additional tools for policymakers.
While we considered settings where medical treatment has a fixed price, as in the motivating study, where treatment availability and associated medical costs are administered solely by the Tanzanian government, we note that in practice, price may be negotiated between suppliers and the purchasing government.
Reframing the research question could help determine the price threshold that would allow for feasible implementation of potential treatments by policymakers.
While the choice among SRT, DLY, and DST scenarios should be guided by research and program needs, our method could also be applied to assess when the choice of estimands leads to meaningful differences and quantify them.}

Our proposed methodology allows for violations of the proportional hazards assumption. 
The current methods were developed in the setting of time-independent covariates, and ongoing research aims to extend our methodology to allow for time-varying covariates, time-varying costs, and the estimation of cost-effectiveness over the full life course.
This paper considers settings where medical costs per time unit are independent from the time of initiation of treatment and from the time-to-event outcome, as in the motivating study, where treatment availability and the associated medical expenses are administered by the Tanzanian government alone while outcomes and times of initiation of treatment are determined by the patients and their providers.
\textcolor{black}{Moreover, we note that survival may depend on the time of treatment initiation. 
In our motivating example, we adjusted for the time of initiation by including it as a term in the Cox model and assumed that there was no model misspecification with respect to this or other model variables.
As always, misspecification of the Cox model could lead to biased RMST estimates.
It is important to note that the methods presented here immediately generalize to allow more flexible models, for example, allowing non-linear relationships between time of initiation with the hazard, for example, by using splines, allowing interactions of time of initiation with other covariates, or by stratifying by time of initiation, thereby eliminating the need to model it. Sensitivity analyses can also be used to compare results of different model assumptions.
}

The ICER and INB are functions of the restricted mean survival time. 
The maximum value of the restricted mean is limited by the period during which data are observed. The difference between restricted means is not necessarily a linear function of the time duration to which the RMST refers. 
It is possible that the difference in restricted means between treatments in a short period of time, as considered by us in a 6 year period, is small, but over 20 year period, the 2nd line ARV2 may be much more cost-effective, and the difference between the restricted means could be much greater. 
A future research topic will be to extrapolate RMSTs over time periods beyond the observed data, as health economists typically do [\cite{WB}].
Alternatively, we could consider quality-adjusted survival times [\cite{ZT,ZT2}] to construct estimators of quality-adjusted ICER and INB based on the relationship between net cost and changes in health outcomes. 
We will extend the methods in this paper to consider quality-adjusted ICER and INB in a future research.
The rich literature on causal inference includes approaches such as $g$-estimation of marginal means.
After identifying counterfactual probabilities for time-to-event outcomes, these estimation methods compute a weighted average of the parameter of interest along the causal pathway  [\cite{Robins1986,rubin2005,Wen2021}].
Comparisons between our method and causal inference methods for cost-effectiveness measures in survival analysis will be a topic for future research.
In addition, to estimate the variance of the ICER and INB estimators, one could consider using bootstrap methods, which is applicable in survival analysis [\cite{Xu2020,yang2024soft}].


\textcolor{black}{Our study was motivated by a study of the comparative effectiveness of switching from first line (ARV1) to second line (ARV2) antiretroviral drugs in relation to all-cause mortality among patients attending HIV clinics in Dar es Salaam, Tanzania, between 2004 to 2012.
Usually, researchers are interested in comparing two treatments; multi-arm trials have less power than two-arm trials.  
However, multi-arm trials do occur, for example, Boland and colleagues reviewed 14 randomized controlled trials making two or three way comparisons of six thrombolytic treatments [\cite{boland2003early}].
The methods presented here immediately
allow for pairwise comparison among multiple treatments.}
\textcolor{black}{Lastly, we note that if all, or nearly all, patients survive beyond the maximum observation time, there would be little information about the survival function and we would not be able to estimate the RMST.
Still, in epidemiological studies and clinical trials, investigators typically choose reasonable follow-up periods according to the survival endpoints and study populations of interest, so that it is not likely to observe no events at all.
}

In summary, this paper provides methodology for estimation and inference for measures of cost-effectiveness allowing for a delay in treatment initiation. 
We also identified further directions in which our methodology can be extended in order to provide an even more helpful toolkit for policy makers.

\section*{Acknowledgments}

This work was conducted with support from the National Institute of Health grants R01AI112339 and DP1ES025459.



\subsection*{Financial disclosure}

None reported.

\subsection*{Conflict of interest}

The authors declare no potential conflict of interests.

\section*{Data availability}

The SAS program for the proposed method is available on GitHub https://github.com/tgh1122334/ ICER/tree/main.

\section*{Supporting information}

Web Appendices referenced in Sections 2.3, 2.4, 2.6, 3, 4, and 5 are available with this article in the Journal of the Royal Statistical Society - Series C website.

\appendix

\section*{A. Definitions for scenario STRT\label{app1a}}

\[
G_j^{(0)}(\boldsymbol{\chi},\boldsymbol{\hat\beta},t)=\frac{1}{n_j}\sum_{i=1}^{n_j}I_{\{T_{ij}>t\}}\exp(\boldsymbol{\hat\beta}^T \boldsymbol{X}_{ij}),\; j=1,\ldots,J. \;\;\;
\]
\[
\boldsymbol{G}_j^{(1)}(\boldsymbol{\chi},\boldsymbol{\hat\beta},t)=\frac{1}{n_j}\sum_{i=1}^{n_j}I_{\{T_{ij}>t\}}\boldsymbol{X}_{ij}\exp(\boldsymbol{\hat\beta}^T \boldsymbol{X}_{ij}).
\]

\[
{\boldsymbol H}_j(\boldsymbol{\chi},\boldsymbol{\hat\beta},t)=\frac{1}{n_j}\sum_{p:T_{j(p)}\leq t}\frac{\boldsymbol{G}_j^{(1)}(\boldsymbol{\chi},\boldsymbol{\hat\beta},T_{j(p)})}{G_j^{(0)}(\boldsymbol{\chi},\boldsymbol{\hat\beta},T_{j(p)})^2}. \quad\quad\quad\quad
{\boldsymbol H}_j^{(r)}(\boldsymbol{\chi},\boldsymbol{\hat\beta},t)=\frac{1}{n_j}\sum_{p:r\leq T_{j(p)}\leq t}\frac{\boldsymbol{G}_j^{(1)}( \boldsymbol{\chi},\boldsymbol{\hat\beta},T_{j(p)})}{G_j^{(0)}(\boldsymbol{\chi},\boldsymbol{\hat\beta},T_{j(p)})^2}.
\]

\[
\widehat{V}_j(\boldsymbol{\chi},t)=\frac{1}{n_j}\sum_{p:T_{j(p)}\leq t}\frac{1}{G_j^{(0)}(\boldsymbol{\chi},\boldsymbol{\hat{\beta}},T_{j(p)})^2}. \quad\quad\quad\quad
\widehat{V}_{j}^{(r)}(\boldsymbol{\chi},t)=\frac{1}{n_j}\sum_{p:r\leq T_{j(p)}\leq t}\frac{1}{G_j^{(0)}(\boldsymbol{\chi},\boldsymbol{\hat{\beta}},T_{j(p)})^2}.
\]

\[
\widehat\Gamma_j(\boldsymbol{\chi},t)=\frac{1}{n}\sum_{k=1}^J\sum_{i=1}^{n_k}\exp(\boldsymbol{\hat\beta}^T \boldsymbol{X}_{ik})\widehat S_j(t| \boldsymbol{X}_{ik}). \quad\quad\quad\quad
\widehat\Gamma_j^{(r)}(\boldsymbol{\chi},t)=\frac{1}{n}\sum_{k=1}^J\sum_{i=1}^{n_k}\exp(\boldsymbol{\hat\beta}^T \boldsymbol{X}_{ik})\widehat S_j^{(r)}(t| \boldsymbol{X}_{ik}).
\]
\[
{\Omega}_j^{(r)}(\boldsymbol{\chi})=\sum_{p:r\leq T_{j(p)}\leq \eta}\sum_{q:r\leq T_{j(q)}\leq \eta}\widehat{V}_j^{(r)}(\boldsymbol{\chi},T_{j(p-1)}\wedge T_{j(q-1)})\widehat{\Gamma}_j^{(r)}(\boldsymbol{\chi},T_{j(p-1)})\widehat{\Gamma}_j^{(r)}(\boldsymbol{\chi},T_{j(q-1)})D_{jp}D_{jq},
\]
where $a\wedge b=max\{a,b\}$,  and $D_{jp}=T_{j(p)}-T_{j(p-1)}$.
\[
{\boldsymbol{\Psi}}_j^{(r)}(\boldsymbol{\chi})=\sum_{p:r\leq T_{j(p)}\leq \eta}\left[\widehat\Gamma_j^{(r)}(T_{j(p-1)})\boldsymbol{H}_j^{(r)}(\boldsymbol{\boldsymbol{\chi},\hat\beta},T_{j(p-1)})-
\widehat{\Lambda}_{0j}(T_{j(p-1)})\boldsymbol{\widehat{\Phi}}_j^{(r)}(\boldsymbol{\chi},T_{j(p-1)})\right]D_{jp},
\]
where  $\widehat{\Lambda}_{0j}(\cdot)$ is  in \eqref{LAMBDA}, 
and
$
\boldsymbol{\widehat{\Phi}}_j^{(r)}(\boldsymbol{\chi},t)=\frac{1}{n}\sum_{k=1}^J\sum_{i=1}^{n_k}\boldsymbol{X}_{ik}\exp(\boldsymbol{\hat\beta}^T \boldsymbol{X}_{ik})\widehat S_j^{(r)}(t| \boldsymbol{X}_{ik}).
$
\[
Cov(\bar\mu_{1}^{STRT}({\boldsymbol{\chi}},r),\bar\mu_{j}^{STRT}({\boldsymbol{\chi}},r))=
n_j^{-1}{\boldsymbol{{\Psi}}}_1^{(r)}(\boldsymbol{\chi})^{T}{\boldsymbol{\widehat{\Sigma}}}^{-1}
{\boldsymbol{{\Psi}}}_j^{(r)}(\boldsymbol{\chi}).
\]


\section*{B. Definitions for scenarios DLY and DST\label{app1b}}
Using the quantities defined in Appendix A, we further define the following quantities.
\[
{\Omega}_j(\boldsymbol{\chi})=\sum_{p:T_{j(p)}\leq \eta}\sum_{q:T_{j(q)}\leq \eta}\widehat{V}_j(\boldsymbol{\chi},T_{j(p-1)}\wedge T_{j(q-1)})\widehat{\Gamma}_j(\boldsymbol{\chi},T_{j(p-1)})\widehat{\Gamma}_j(\boldsymbol{\chi},T_{j(q-1)})D_{jp}D_{jq}, \;j=1,\ldots, J,
\]
\[
{\boldsymbol{{\Psi}}}_1(\boldsymbol{\chi})=\sum_{p:T_{j(p)}\le \eta}\left[\widehat\Gamma_1(\boldsymbol{\chi},T_{j(p-1)})\boldsymbol{H}_1(\boldsymbol{\chi},\hat\beta, T_{j(p-1)})-\widehat{\Lambda}_{01}(T_{j(p-1)})\widehat{\boldsymbol{\Phi}}_1(\boldsymbol{\chi},T_{j(p-1)})\right]D_{1p},
\]
where  $
\boldsymbol{\widehat{\Phi}}_1(\boldsymbol{\chi},t)=\frac{1}{n}\sum_{k=1}^J\sum_{i=1}^{n_k}\boldsymbol{X}_{ik}\exp(\boldsymbol{\hat\beta}^T \boldsymbol{X}_{ik})\widehat S_1(t| \boldsymbol{X}_{ik})
$, and
$\widehat{\Lambda}_{0j}(\cdot)$ is in \eqref{LAMBDA}. 
\[
\widetilde{\Omega}_1^{(a)}(\boldsymbol{\chi})=\sum_{p:T_{j(p)}< a}\sum_{q:T_{1(q)}< a}\widehat{V}_1(\boldsymbol{\chi},T_{1(p-1)}\wedge T_{1(q-1)})\widehat{\Gamma}_1(\boldsymbol{\chi},T_{1(p-1)})\widehat{\Gamma}_1(\boldsymbol{\chi},T_{1(q-1)})D_{1p}D_{1q}.
\]
\[
\widetilde{\Omega}_j^{(a)}(\boldsymbol{\chi})=\sum_{p:a\leq T_{j(p)}\leq \eta}\sum_{q:a\leq T_{j(q)}\leq \eta}\widehat{V}_j^{(a)}(\boldsymbol{\chi},T_{j(p-1)}\wedge T_{j(q-1)})\widehat{\widetilde\Gamma}_j^{(a)}(\boldsymbol{\chi},T_{j(p-1)})\widehat{\widetilde\Gamma}_j^{(a)}(\boldsymbol{\chi},T_{j(q-1)})D_{jp}D_{jq},
\]
where  $
\widehat{\widetilde \Gamma_j}^{(a)}(\boldsymbol{\chi},t)=\frac{1}{n}\sum_{k=1}^J\sum_{i=1}^{n_k}\exp(\boldsymbol{\hat\beta}^T \boldsymbol{X}_{ik})\widehat{\widetilde S}_j^{(a)}(t|\boldsymbol{X}_{ik}).
$ 
\[
\widetilde{\Omega}_1^{(\delta_{lj},\delta_{hj})}(\boldsymbol{\chi})=\sum_{\substack{
		p: T_{j(p)}\leq \min(\delta_{lj},\delta_{hj})\\ q: T_{j(q)}\leq \min(\delta_{lj},\delta_{hj})}}
\widehat{V}_1(\boldsymbol{\chi},T_{j(p-1)}\wedge T_{j(q-1)})\widehat{\Gamma}_1(\boldsymbol{\chi},T_{j(p-1)})\widehat{\Gamma}_1(\boldsymbol{\chi},T_{j(q-1)})D_{jp}D_{jq}.
\]
\[
\widetilde{\Omega}_j^{(\delta_{lj},\delta_{hj})}(\boldsymbol{\chi})=\sum_{\substack{p:\max(\delta_{lj},\delta_{hj})\leq T_{j(p)}\leq\eta \\ q: \max(\delta_{lj},\delta_{hj})\leq T_{j(q)}\leq \eta}}\widehat{V}_j^{(a)}(\boldsymbol{\chi},T_{j(p-1)}\wedge T_{j(q-1)})\widehat{\widetilde\Gamma}_j^{(a)}(\boldsymbol{\chi},T_{j(p-1)})\widehat{\widetilde\Gamma}_j^{(a)}(\boldsymbol{\chi},T_{j(q-1)})D_{jp}D_{jq}.
\]
\[
{\boldsymbol{\widetilde{\Psi}}}^{(a)}_1(\boldsymbol{\chi})=\sum_{p:T_{j(p)}< a}\left[\widehat\Gamma_1(\boldsymbol{\chi},T_{j(p-1)})\boldsymbol{H}_1(\boldsymbol{\chi},\hat\beta, T_{j(p-1)})-\widehat{\Lambda}_{01}(T_{j(p-1)})\widehat{\boldsymbol{\Phi}}_1(\boldsymbol{\chi},T_{j(p-1)})\right]D_{1p}.
\]
\[
{\boldsymbol{\widetilde{\Psi}}}^{(\delta_{lj},\delta_{hj})}_1(\boldsymbol{\chi})=\sum_{p:T_{j(p)}< \min(\delta_{lj},\delta_{hj})}\left[\widehat\Gamma_1(\boldsymbol{\chi},T_{j(p-1)})\boldsymbol{H}_1(\boldsymbol{\chi},\boldsymbol{\hat\beta}, T_{j(p-1)})-\widehat{\Lambda}_{01}(T_{j(p-1)})\widehat{\boldsymbol{\Phi}}_1(\boldsymbol{\chi},T_{j(p-1)})\right]D_{1p}.
\]
\[
{\boldsymbol{\widetilde{\Psi}}}_j^{(a)}(\boldsymbol{\chi})=\sum_{p:a\leq T_{j(p)}\leq \eta}\left[\widehat{\widetilde\Gamma}_j^{(a)}(\boldsymbol{\chi},T_{j(p-1)})\boldsymbol{H}_j^{(a)}(\boldsymbol{\chi},\boldsymbol{\hat\beta}, T_{j(p-1)})-\widehat{\Lambda}_{0j}(T_{j(p-1)})\widehat{\widetilde{\boldsymbol{\Phi}}}_j^{(a)}(\boldsymbol{\chi},T_{j(p-1)})\right]D_{jp}.
\]
where $\widehat{\widetilde{\boldsymbol{\Phi}}}_j^{(a)}(\boldsymbol{\chi},t)=\frac{1}{n}\sum_{k=1}^J\sum_{i=1}^{n_k}\boldsymbol{X}_{ik}\exp(\boldsymbol{\hat\beta}^T \boldsymbol{X}_{ik})\widehat{\widetilde S}_j^{(a)}(t|\boldsymbol{X}_{ik}).$
\[
Cov(\bar\mu_1^{DLY}({\boldsymbol{\chi}},a),\bar\mu_j^{DLY}({\boldsymbol{\chi}},a))=
n_j^{-1}\big[\widetilde\Omega_1^{(a)}+{\boldsymbol{\widetilde{\Psi}}}_1^{T}(\boldsymbol{\chi})
{\boldsymbol{\widehat{\Sigma}}}^{-1}
({\boldsymbol{\widetilde{\Psi}}}_1^{(a)}(\boldsymbol{\chi})+{\boldsymbol{\widetilde{\Psi}}}_j^{(a)}(\boldsymbol{\chi}))\big].
\]
\[
\begin{split}
Cov(\bar\mu_{1}^{DST}(\boldsymbol{\chi},{\boldsymbol{\Delta}}),\bar\mu_{j}^{DST}(\boldsymbol{\chi},{\boldsymbol{\Delta}}))&=
n_j^{-1}(n_j-1)^{-1}\sum_{l\ne h}[\widetilde\Omega_1^{(\delta_{lj},\delta_{hj})}+\widetilde\Omega_j^{(\delta_{lj},\delta_{hj})}\\
&+{\boldsymbol{\widetilde{\Psi}}}_1^{(\delta_{lj},\delta_{hj})T}(\boldsymbol{\chi})
{\boldsymbol{\widehat{\Sigma}}}^{-1}
{\boldsymbol{\widetilde{\Psi}}}_1^{(\delta_{lj},\delta_{hj})}(\boldsymbol{\chi})+{\boldsymbol{\widetilde{\Psi}}}_j^{(\delta_{lj})T}(\boldsymbol{\chi})
{\boldsymbol{\widehat{\Sigma}}}^{-1}
{\boldsymbol{\widetilde{\Psi}}}_j^{(\delta_{hj})}(\boldsymbol{\chi}) ].
\end{split}
\]

\bibliographystyle{agsm}
\bibliography{Reference.bib}

\newpage
\normalsize

\normalsize

\begin{table}
	\caption{\label{tab1} Simulation study results: Performance of $\widehat{\text{RMST}}$ under scenarios STRT and DLY.}
	\centering
		\scriptsize {
			\begin{tabular}{cccccccccc}
				\hline
				Sample&         &         &     & \multicolumn{2}{c}{\% Relative bias of}   &Coverage    & \multicolumn{2}{c}{\% Relative bias of}      & Coverage      \\
				Size  &Scenario	&Delay (\%)&	HR	& $\widehat{\mu}_1$	& $\widehat{\text{SE}}$(${\mu}_1$)	&probability & $\widehat{\mu}_2$	& $\widehat{\text{SE}}$(${\mu}_2$)	& probability 	\\
				\hline
				1,000 & No delay	&	0\%	&	0.2	& -0.1	&	-2.6	&	0.94	&	0	&	-5.3	&	0.93	\\
				&	            &		&	0.5	& -0.2	&	-2.0	&	0.95	& -0.1	&	-8.5	&	0.93	\\
				&         	&		&	0.8	& -0.1	&	-6.7	&	0.93	& -0.1	&	-2.4	&	0.94	\\
				&         	&		&		&		&		&		&		&		&		\\
				& STRT	    & 10\%	&	0.2	&  0.1	&	-0.7	&	0.95	&  0.6	&	-9.5	&	0.91	\\
				& (r=0.5)	    &		&	0.5	&	0	&	-2.2	&	0.95	&  0.2	&	-6.4	&	0.93	\\
				&	            &		&	0.8	&	0	&	-6.4	&	0.93	&  0.2	&	-4.0	&	0.94	\\
				&	&		&		&		&		&		&		&		&		\\
				&	            & 50\%	&	0.2	& -0.1	&	-0.7	&	0.95	&  0.7	&	-9.6	&	0.90	\\
				&	            &		&	0.5	& -0.2	&	-2.3	&	0.94	&  0.1	&	-6.7	&	0.93	\\
				&	            &		&	0.8	&  0.2	&	-1.2	&	0.94	& -0.1	&	-3.7	&	0.93	\\
				&	&		&		&		&		&		&		&		&		\\
				& DLY	        & 10\%	&	0.2	&  0	&	-8.4	&	0.95	& -0.5  &	-20.9	&	0.90	\\
				& (a=0.5)	    &		&	0.5	& -0.1	&	-8.1	&	0.94	& -0.2	&	-12.0	&	0.93	\\
				&	            &		&	0.8	&	0	&	-12.0	&	0.94	& -0.2	&	-10.6	&	0.94	\\
				&	&		&		&		&		&		&		&		&		\\
				&          	& 50\%	&	0.2	& -0.1	&	-8.4	&	0.94	&	0.5	&	-20.3	&	0.90	\\
				&	            &		&	0.5	& -0.4	&	-7.6	&	0.95	& -0.1	&	-11.3	&	0.93	\\
				&         	&		&	0.8	& -0.5	&	-11.8	&	0.93	& -0.2	&	-10.0	&	0.94	\\
				\multicolumn{10}{c}{ }  \\																	
				10,000 & No delay	&	0\%	&	0.2	&	0	&	-4.6	&	0.95	&	0	&	-5.5	&	0.93	\\
				&	        &		&	0.5	&	0	&	-3.6	&	0.94	&	0	&	-7.8	&	0.93	\\
				&	        &		&	0.8	&	0	&	-5.2	&	0.94	&	0	&	-2.4	&	0.94	\\
				&	&		&		&		&		&		&		&		&		\\
				& STRT	    &	10\%	&	0.2	&	0	&	-3.7	&	0.94	&	0	&	-4.8	&	0.94	\\
				& (r=0.5    )&		    &	0.5	&	0	&	-4.9	&	0.94	&	0	&	-5.5	&	0.93	\\
				&	        &	      	&	0.8	&  -0.2	&	-0.6	&	0.94	&	0	&	-1.6	&	0.93	\\
				&	&		&		&		    &		&		&		&		&		\\
				&	        &	50\%	&	0.2	&  -0.1	&	-3.8	&	0.94	&	0.0	&	-5.3	&	0.93	\\
				&	        &	     	&	0.5	&  -0.3	&	-4.7	&	0.92	&  -0.3	&	-5.5	&	0.92	\\
				&	        &	    	&	0.8	&  -0.4	&	-3.3	&	0.92	&  -0.5	&	-3.4	&	0.90	\\
				&	&		&		&		   &		&		&		&		&		\\
				& DLY	        & 10\%	    &	0.2	&  0	&	-10.6	&	0.94	&	0.1	&   -8.0	&	0.95	\\
				&	(a=0.5)     &		    &	0.5	&  0	&	-8.8	&	0.94	&	0	&	-8.9	&	0.95	\\
				&	            &		    &	0.8	&  0 	&	-9.6	&	0.94	&	0	&	-6.7	&	0.95	\\
				&	&		&		&		&		&		&		&		&		\\
				&	            &	50\%	&	0.2	&  0	&	-10.6	&	0.94	&	0.1	&	-8.0	&	0.95	\\
				&          	&		    &	0.5	&  0    &	-8.8	&	0.94	&	0	&	-8.9	&	0.94    \\
				&	            &		    &	0.8	&  0    &	-9.6	&	0.94	&	0	&	-6.7	&	0.95	\\
				\hline
\multicolumn{10}{l}{Note: Scenario STRT: for patients who survive to a fixed time $r$; Scenario DLY: for a given fixed delay time $a$;}\\
 \multicolumn{10}{l}{Sample size is 10,000; the number of simulation replicates is 1,000; controlling for one covariate generated from a} \\
\multicolumn{10}{l}{Bernoulli distribution with probability for success 0.9. }
			\end{tabular}
		}
\end{table}

\begin{table}
	\caption{\label{tab2} Simulation results: Performance of $\widehat{\text{ICER}}$ and $\widehat{\text{INB}}^*$ under scenarios STRT and DLY.}
	\centering
		\scriptsize{
			\begin{tabular}{cccccccccc}
				\hline
				Sample&         &         &     & \% Relative Bias of & coverage    & \% Relative Bias & coverage   \\
				Size  &Scenario	&Delay (\%)&	HR	&   	$\widehat{\text{ICER}}$	      & probability &       $\widehat{\text{INB}}$	       & probability	\\
				\hline
				1,000   &No delay	&	0\%	&	0.2	& -0.4	&	0.95	&	 0.2	&	0.95	\\
				&    	    & 		&	0.5	& -1.5	&	0.95	&	1.0 	&	0.95	\\
				&	        &	  	&	0.8	& -18.3 &	0.91	&	-0.8   	&	0.95	\\
				&	        &	    &		&		&			&		    &			\\
				&STRT       &	10\%&	0.2	&  -1	&	0.96	&	2.2	&	0.94	\\
				&(r=0.5)    &		&	0.5	& -1.9	&	0.94	&	2.2	    &	0.95	\\
				&	        &		&	0.8	& -25	&	0.91	&  -2.0	    &	0.94	\\
				&	        &		&		&		&	     	&		    &			\\
				&	        &	50\%&	0.2	& -1.3	&	0.96	&	 2.8	&	0.94	\\
				&	        &		&	0.5	& -2.3	&	0.94	&	 3.5	&	0.95	\\
				&	        &		&	0.8	& -3.2	&	0.92	&	 2.0	&	0.96	\\
				&	        &		&		&		&			&		    &			\\
				& DLY	    &	10\%&	0.2	&	1	&	0.96	&	 2.2	&	0.95	\\
				&(a=0.5)    &		&	0.5	&	1.9	&	0.94	&	 2.2	&	0.94	\\
				&	        &	    &	0.8	&	24.1&	0.91	&	 -2.0  	&	0.94	\\
				&	        &		&		&		&			&		    &			\\
				&	        &	50\%&	0.2	& -1.3	&	0.96	&	 2.7	&	0.94	\\
				&	        &		&	0.5	& -2.2	&	0.95	&	 3.3	&	0.95	\\
				&	        &		&	0.8	& 28.9	&	0.92	&	-3.2    &	0.94	\\
				&	        &		&		&		&	    	&		    &			\\
				10,000  &No delay	&	0\%	&	0.2	&	0	&	0.95	&	0	    &	0.95	\\
				&	        &		&	0.5	& -0.1	&	0.95	&	0.1  	&	0.95	\\
				&	        &		&	0.8	& -1.5	&	0.96	&	0.7	    &	0.96	\\
				&	        &		&		&		&			&		    &	    	\\
				& STRT	    &	10\%&	0.2	& -0.1	&	0.96	&	0.2 	&	0.95	\\
				&(r=0.5)    &		&	0.5	& -0.1	&	0.95	&	0	    &	0.94	\\
				&	        &		&	0.8	&  0	&	0.96	&	1.3 	&	0.97	\\
				&	        &		&		&		&			&		    &			\\
				&	        &	50\%&	0.2	&	0.1	&	0.95	&	-0.3	&	0.95	\\
				&    	    &		&	0.5	&	0.2	&	0.95	&	-0.2	&	0.95	\\
				&	        &		&	0.8	&	0.8	&	0.95	&	-0.2	&	0.95	\\
				&	        &		&		&		&			&		    &			\\
				& DLY	    &	10\%&	0.2	& -0.1	&	0.94	&	0.1	    &	0.95	\\
				&(a=0.5)    &		&	0.5	& -0.1	&	0.95	&	-0.2	&	0.94	\\
				&	        &		&	0.8	& -1.4	&	0.94	&	-0.2	&	0.95	\\
				&	        &		&		&		&			&		    &			\\
				&	        &	50\%&	0.2	& -0.1	&	0.93	&	0.1	    &	0.94	\\
				&	        &		&	0.5	& -0.1	&	0.94	&	-0.2    &	0.94	\\
				&	        &		&	0.8	& -1.4	&	0.94	&	-0.2	&	0.95	\\
				\hline
				\multicolumn{10}{l}{* Assuming that willingness-to-pay, $\lambda$, is equal to \$1,352.} \\
\multicolumn{10}{l}{Note: Scenario STRT: for patients who survive to a fixed time $r$; Scenario DLY: for a given fixed delay time $a$;}\\
\multicolumn{10}{l}{Sample size is 10,000; the number of simulation replicates is 1,000;  controlling for  one covariate generated from a} \\
\multicolumn{10}{l}{ Bernoulli distribution with probability for success 0.9. }
			\end{tabular}
		}
\end{table}

\begin{table}
	\caption{\label{tab3} Simulation study results: Performance of our $\widehat{\text{RMST}}$, $\widehat{\text{ICER}}$ and $\widehat{\text{INB}}^*$  in scenario DST (i.e., for a given distribution of delay time).}
	\centering
		\scriptsize{
			\begin{tabular}{cccccc}\hline
				Delay time &   & \% Relative bias of & Coverage    &\% Relative bias of & Coverage     \\
				distribution & HR&	$\widehat{\mu_1}$  & probability &$\widehat{\mu_2}$	  & probability	\\
				\hline
				Uniform (0,1)&	0.2	& 0.1	&	0.93	&	0	&	0.98	\\
				&	0.5	&  0	&	0.95	& 0.2	&	0.97	\\
				&	0.8	&	0	&	0.94	&	0	&	0.97	\\
				&		&		&			&		&			\\
				Option 2**	 &	0.2	&	0.1	&	0.93	&	0.1	&	0.98	\\
				&	0.5 &	0	&	0.95	&	0	&	0.96	\\
				&	0.8 &	0	&	0.95	&	0	&	0.97	\\
				&		&		&			&		&			\\
				Option 3**	 &	0.2	&	0.1	&	0.94	&	0.1	&	0.98	\\
				&	0.5	&	0	&	0.94	&	0	&	0.99	\\
				&	0.8	&	0	&	0.93	&	0	&	0.97	\\
				\hline
				Delay time &   &\% Relative bias of      & Coverage    &\% Relative bias of     & Coverage     \\
				distribution&HR	&$\widehat{\text{ICER}}$  & probability & $\widehat{\text{INB}}$ & probability	\\
				\hline
				Uniform (0,1)	&	0.2	&	0	&	0.95	& -0.3	&	0.95	\\
				&	0.5	& -0.4	&	0.95	&  1.6	&	0.94	\\
				&	0.8	& -1.4	&	0.95	& -0.4	&	0.96	\\
				&		&		&			&		&			\\
				Option 2**	    &	0.2	&	0.0	&	0.94	& -0.2	&	0.95	\\
				&	0.5	& -0.1	&	0.94	&	0	&	0.95	\\
				&	0.8	& -1.3	&	0.95	& -0.3	&	0.97	\\
				&		&		&	    	&		&			\\
				Option 3**	    &	0.2	&	0	&	0.95	&  0	&	0.96	\\
				&	0.5	&	0	&	0.95	& -0.5	&	0.95	\\
				&	0.8	& -1.0	&	0.96	& 0.4	&	0.96	\\
				\hline
				\multicolumn{6}{l}{*Assuming that willingness-to-pay, $\xi$, is equal to \$1,352} \\
				\multicolumn{6}{l}{**Delay time distributions for Options 2 and 3 are presented in Supplementary Figure 1 }\\
\multicolumn{6}{l}{Note: Sample size is 10,000; the number of simulation replicates is 1,000;  controlling for  one covariate } \\
\multicolumn{6}{l}{  generated from a Bernoulli distribution with probability for success 0.9. }
			\end{tabular}
		}
\end{table}

\begin{table}
	\caption{\label{rdata} Cost-effectiveness analysis of ARV2 switch among ARV2 eligible patients, Dar es Salaam, Tanzania (131/1,455=deaths/patients)*.}
	\centering
		\tiny{
			\begin{tabular}{cccccccc}
				\hline
				Follow-up  & Fixed delay & Covariate    &          &  $\bar\mu_1$& $\bar\mu_2$ &               &               \\
				period (years)& time (a)     & distribution & Scenario & (years)    & (years)&$\widehat{ICER}$ (95\% CI) & $\widehat{INB}$*** (95\% CI)   \\
				\hline
				5	&	0	&	Observed	&	No delay**	&	4.47	&	4.80	&	3,268 (1,415, 5,122)	&	-627 (-858, -397)	\\
				&	0.9	&		&	STRT	&	3.77	&	4.06	&	3,151 (1,569, 4,733)	&	-517 (-697, -338)	\\
				&	0.9	&		&	DLY	&	4.46	&	4.75	&	2,984 (1,625, 4,343)	&	-474 (-639, -309)	\\
				&	-	&		&	DST	&	4.46	&	4.59	&	4,868 (-2,136, 11,871)	&	-606 (-973, -239)	\\
				&		&		&		&		&		&		&		\\
				5.5	&	0	&	Observed	&	No delay**	&	4.89	&	5.27	&	3,110 (1,331, 4,888)	&	-665 (-934, -395)	\\
				&	0.9	&		&	STRT	&	4.21	&	4.53	&	3,103 (1,461, 4,743)	&	-571 (-785, -357)	\\
				&	0.9	&		&	DLY	&	4.88	&	5.21	&	2,269 (1,528, 4,410)	&	-557 (-724, -330)	\\
				&	-	&		&	DST	&	4.89	&	5.04	&	4,485 (-1,831, 10,800)	&	-657 (-1,092, -222)	\\
				&		&		&		&		&		&		&		\\
				6	&	0	&	Observed	&	No delay**	&	5.29	&	5.46	&	2,977 (1,265, 4,688)	&	-700 (-1,009, -390)	\\
				&	0.9	&		&	STRT	&	4.64	&	5.00	&	3,045 (1,388, 4,701)	&	-621 (-868, -375)	\\
				&	0.9	&		&	DLY	&	5.29	&	5.65	&	2,939 (1,459, 4,419)	&	-577 (-803, -350)	\\
				&	-	&		&	DST	&	5.29	&	5.49	&	4,198 (-1,646, 10,043)	&	-706 (-1,212, -199)	\\
				&		&		&		&		&		&		&		\\
				6	&	0	&	Median values	&	No delay**	&	5.73	&	5.97	&	7,514 (2,562, 12,466)	&	-987 (-1,206, -768)	\\
				&	0.9	&		&	STRT	&	4.93	&	5.13	&	5,604 (2,408, 8,801)	&	-855 (-1,005, -705)	\\
				&	0.9	&		&	DLY	&	5.73	&	5.95	&	5,604 (2,408, 8,801)	&	-855 (-1,005, -705)	\\
				&	-	&		&	DST	&	5.73	&	5.76	&	11,847 (151, 23,544)	&	-954 (-1,077, -831)	\\
				&		&		&		&		&		&		&		\\
				6	&	0	&	Sick patients****	&	No delay**	&	3.88	&	5.06	&	1,007 (307, 1,708)	&	421 (-713, 1,557)	\\
				&	0.9	&		&	STRT	&	3.62	&	4.49	&	1,215 (105, 2,326)	&	120 (-962, 1,202)	\\
				&	0.9	&		&	DLY	&	3.87	&	4.59	&	1,234 (311, 2,158)	&	85 (-654, 824)	\\
				&	-	&		&	DST	&	3.87	&	4.50	&	1,334 (-78, 2,747)	&	12 (-946, 969)	\\
				\hline
\multicolumn{8}{l}{Scenario STRT: for patients who suvive to a fixed time $a$; Scenario DLY: for a given fixed delay time $a$;}\\
\multicolumn{8}{l}{Scenario DST: for the observed distribution of delay time; $\bar{\mu}_1$: estimated RMST for ARV1; $\bar{\mu}_2$: estimated RMST for ARV2.}\\
\multicolumn{8}{l}{* Adjusted for gender, tuberculosis (TB) history (yes, no), HIV stage at second line eligibility, marital}\\
\multicolumn{8}{l}{status, age at eligibility ($<$30, 30-$<$40, 40-$<$50, 50+ years), BMI ($<$18.5, 18.5-$<$25, 25-$<$30, 30+ kg/$m^2$), }\\
\multicolumn{8}{l}{hemoglobin ($<$7.5, 7.5-$<$10, 10+ g/dl), CD4 ($<$50, 50-$<$100, 100-$<$200, 200+ cells/$mm^3$), use of cotrimoxazole (yes, no),}\\
\multicolumn{8}{l}{facility level (hospital, health center, dispensary), district of Dar es Salaam, visit adherence (proportion of days late}\\
\multicolumn{8}{l}{for scheduled clinic visit), delay times (time since ARV2 eligibility to the ARV2 initiation) and treatment group (0,1)}\\
\multicolumn{8}{l}{** Scenarios STRT \& DLY give the same results when $a=0$ (no delay)}\\
\multicolumn{8}{l}{*** Calculated assuming that willingness-to-pay, $\xi$, equals \$1,352}\\
\multicolumn{8}{l}{**** Females with BMI less than 18.5, who were 50+ years old at eligibility, with all other covariates as observed in the population}\\
				
			\end{tabular}
		}
\end{table}

\end{document}


\begin{center}
\Large{Supplementary Material for ``Statistical methods for cost-effectiveness analysis of left-truncated censored survival data with treatment delays"}

{\normalsize Polyna Khudyakov, Li Xu, Ce Yang, Donna Spiegelman, Molin Wang}
\end{center}

\vspace{10mm}
\section*{{Supplementary Appendix 1. Derivation of the asymptotic variance for scenario  ${STRT}$}}

The estimator of $S_{j}^{(a)}(t|{\boldsymbol{X}})$ is 
$\widehat S_{j}^{(a)}(t|{\boldsymbol{X}})=\exp\{-e^{{\hat{\boldsymbol{\beta}}}^T{\boldsymbol{X}}}(\widehat\Lambda_{0j}(t)-\widehat\Lambda_{0j}(a))\}$, where $\widehat{\Lambda}_{0j}(t)$ is given by (1) in the manuscript.
Thus, we obtain
\[
\begin{split}
-\sqrt{n_j}(\widehat{S}_{j}^{(a)}(t|{\boldsymbol{X}})-S_{j}^{(a)}(t|{\boldsymbol{X}}))&=e^{{\boldsymbol{\hat\beta}}^T{\boldsymbol{X}}}
S_{j}^{(a)}(t|{\boldsymbol{X}})W_{aj}(\boldsymbol{\chi},t)-\sqrt{n_j}\Big[e^{{\boldsymbol{\beta}}^T{\boldsymbol{X}}}
S_{j}^{(a)}(t|{\boldsymbol{X}}){\boldsymbol{H}}_j^{(a)}({\boldsymbol{\chi}},{\boldsymbol{\beta}},t)\\
&-e^{{\boldsymbol{\beta}}^T{\boldsymbol{X}}}(\Lambda_{0j}(t)-\Lambda_{0j}(a))
S_{j}^{(a)}(t|{\boldsymbol{X}}){\boldsymbol{X}}\Big]^T({\boldsymbol{\hat\beta}}
-{\boldsymbol{\beta}})+o_p(1),\\
\end{split}
\]
where $W_{aj}(\boldsymbol{\chi},t)$ is a martingale
\[
W_{aj}(\boldsymbol{\chi},t)=\int_a^t\frac{{n_j}^{1/2}d\bar M(s)}{\sum_{i=1}^{n_j} Y_{ij}(s) e^{{\boldsymbol{\beta}}{_0^T\boldsymbol{X}_{ij}}}},
\]
converging weakly to an independent increment Gaussian process $W_{aj}^{\infty}$, and $\boldsymbol{H}_j^{(a)}({\boldsymbol{\chi}},\boldsymbol{\beta},t)$ is defined in Appendix A of the manuscript.
The limiting variance function of $W_{aj}^{\infty}$ may be estimated by $\hat{V}_j^{(a)}(\boldsymbol{\chi},t)$, which is 
defined in Appendix A of the manuscript. The asymptotic properties of $\bar{\mu}_{j}^{STRT}$ and variance estimator   $\widehat{Var}(\bar{\mu}_{j}^{STRT}({\boldsymbol{\chi}},a))$ in (2) of the manuscript then follow.

\newpage
\section*{Supplementary Appendix 2. Derivation of the asymptotic variance for  scenario ${DLY}$}

The RMST for group 1 is given by
$\widehat\mu_{1}^{DLY}({\boldsymbol{X}},a)=\int_0^\eta \widehat S_1(t|{\boldsymbol{X}})dt$, with $S_1(t|{\boldsymbol{X}})=\exp\{-e^{{\boldsymbol{\beta}}^T{\boldsymbol{X}}}\Lambda_{01}(t)\}$.
The asymptotic distribution of RMST for subjects that switched to Treatment $j$, $\widehat{\mu}_{j}^{DLY}({\boldsymbol{X}},a)$, can be derived as the distribution of a sum of $\widehat\mu_{1}({\boldsymbol{X}},a)$  up to $a$, and of a second term defined as
\[\widetilde S_{j}^{(a)}(t|X)=S_1(a|{\boldsymbol{X}})S_j(t|T>a,{\boldsymbol{X}})=\exp\{-e^{{\boldsymbol{\beta}}^T{\boldsymbol{X}}}
\left[\Lambda_{01}(a)-\Lambda_{0j}(a)+\Lambda_{0j}(t)\right]\}.\] 
We then have
\begin{align*}&-\sqrt{n_j}(\widehat{\widetilde S}_{j}^{(a)}(t|{\boldsymbol{X}})
-\widetilde S_{j}^{(a)}(t|{\boldsymbol{X}}))
=e^{{\boldsymbol{\beta}}^T{\boldsymbol{X}}}\Big{\{}\widetilde S_{j}^{(a)}(t|{\boldsymbol{X}})
\left[W_{a1}({\boldsymbol{\chi}},a)+W_{aj}({\boldsymbol{\chi}},t)\right]
\\
&-\sqrt{n_j}
[\widetilde S_{j}^{(a)}(t|{\boldsymbol{X}})({\boldsymbol{H}}_1({\boldsymbol{\chi}},\beta,a)+{\boldsymbol{H}}_j^{(a)}({\boldsymbol{\chi}},\beta,t))
\\&-(\Lambda_{01}(a)-\Lambda_{0j}(a)+\Lambda_{0j}(t))\widetilde S_{j}^{(a)}(t|{\boldsymbol{X}}){\boldsymbol{X}}]^T
({\boldsymbol{\hat\beta}}-{\boldsymbol{\beta}})\Big{\}}\small{+o_p(1)},
\end{align*} where $W_{aj}$ is defined in   Web Supplementary Appendix 1 above and ${\boldsymbol{H}}_1$ and ${\boldsymbol{H}}_j^{(a)}$ are defined in Appendix A of the manuscript. 
The variance estimator in Section 2.4 of the manuscript then follows.

\newpage
\textcolor{black}{\section*{Supplementary Appendix 3. The formula for $ICER^{STRT}$, $ICER^{DLY}$, and $ICER^{DST}$}}
\label{A2}


Under scenario STRT, for covariate distribution $\boldsymbol{\chi}$, the ICER is estimated by
\begin{align*}
\widehat{ICER}_j^{STRT}({\boldsymbol{\chi}}, r) &= \frac{c_j \bar\mu_j^{STRT}({\boldsymbol{\chi}}, r)
- c_1 \bar\mu_1^{STRT}({\boldsymbol{\chi}}, r)}
{\bar\mu_j^{STRT}({\boldsymbol{\chi}}, r) - \bar\mu_1^{STRT}({\boldsymbol{\chi}}, r)} \\
&= \frac{c_j \bar\mu_j^{STRT,2}({\boldsymbol{\chi}}, r)
- c_1 \bar\mu_1^{STRT,2}({\boldsymbol{\chi}}, r)}
{\bar\mu_j^{STRT,2}({\boldsymbol{\chi}}, r) - \bar\mu_1^{STRT,2}({\boldsymbol{\chi}}, r)}.
\end{align*}
Under scenario DLY,
{\footnotesize
\begin{align*}
&\widehat{ICER}_j^{DLY}({\boldsymbol{\chi}}, a) = \\
&\frac{\frac{1}{n} \sum_{k=1}^{J} \sum_{i=1}^{n_k} \left\{ c_1 \int_0^a S_1(t|{\boldsymbol x})dt + c_j \frac{S_1(a|{\boldsymbol x})} {S_j(a|{\boldsymbol x})} \int_a^\eta S_j(t|{\boldsymbol x}) dt \right\} - \frac{1}{n} \sum_{k=1}^{J} \sum_{i=1}^{n_k} c_1 \left\{ \int_0^a S_1(t|{\boldsymbol x}) dt + \frac{S_1(a|{\boldsymbol x})} {S_1(a|{\boldsymbol x})} \int_a^\eta S_1(t|{\boldsymbol x})dt) \right\} } 
{\frac{1}{n} \sum_{k=1}^{J} \sum_{i=1}^{n_k} \left\{ \int_0^a S_1 (t|{\boldsymbol x}) dt + \frac{S_1(a|{\boldsymbol x})}{S_j(a|{\boldsymbol x})} \int_a^\eta S_j(t|{\boldsymbol x}) dt \right \} - \frac{1}{n} \sum_{k=1}^{J} \sum_{i=1}^{n_k} \left\{ \int_0^a S_1(t|{\boldsymbol x}) dt + \frac{S_1(a|{\boldsymbol x})}{S_1(a|{\boldsymbol x})} \int_a^\eta S_1(t|{\boldsymbol x}) dt \right \}} \\
&= \frac{\frac{1}{n} \sum_{k=1}^{J} \sum_{i=1}^{n_k} \left\{ c_j \frac{S_1(a|{\boldsymbol x})} {S_j(a|{\boldsymbol x})} \int_a^\eta S_j(t|{\boldsymbol x}) dt \right\} - \frac{1}{n} \sum_{k=1}^{J} \sum_{i=1}^{n_k} c_1 \left\{ \frac{S_1(a|{\boldsymbol x})} {S_1(a|{\boldsymbol x})} \int_a^\eta S_1(t|{\boldsymbol x})dt) \right\} } 
{\frac{1}{n} \sum_{k=1}^{J} \sum_{i=1}^{n_k} \left\{ \frac{S_1(a|{\boldsymbol x})}{S_j(a|{\boldsymbol x})} \int_a^\eta S_j(t|{\boldsymbol x}) dt \right \} - \frac{1}{n} \sum_{k=1}^{J} \sum_{i=1}^{n_k} \left\{ \frac{S_1(a|{\boldsymbol x})}{S_1(a|{\boldsymbol x})} \int_a^\eta S_1(t|{\boldsymbol x}) dt \right \}} \\
&= \frac{c_j \bar\mu_{j}^{DLY, 2}({\boldsymbol{\chi}}, a) - c_1 \bar\mu_{1}^{DLY, 2}({\boldsymbol{\chi}}, a)}
{\bar\mu_{j}^{DLY, 2}({\boldsymbol{\chi}}, a) - \bar\mu_{1}^{DLY, 2}({\boldsymbol{\chi}}, a)}.
\end{align*}
}
Similarly, under scenario DST,
{\footnotesize
\begin{align*}
&\widehat{ICER}_j^{DST}({\boldsymbol{\chi}}, \boldsymbol{\Delta}) = \frac{c_j \bar\mu_{j}^{DST, 2}({\boldsymbol{\chi}}, \boldsymbol{\Delta}) - c_1 \bar\mu_{1}^{DST, 2}({\boldsymbol{\chi}}, \boldsymbol{\Delta})} {\bar\mu_{j}^{DST, 2}({\boldsymbol{\chi}}, \boldsymbol{\Delta}) - \bar\mu_{1}^{DST, 2}({\boldsymbol{\chi}}, \boldsymbol{\Delta})}.
\end{align*}
}

\newpage
Regarding the relationship between $ICER^{STRT}$ and $ICER^{DLY}$, and $INB^{STRT}$ and $INB^{DLY}$ when $\boldsymbol X=\boldsymbol x$,
\vspace{0.2in}
\[
\begin{split}
&ICER^{DLY}({\boldsymbol x},a) \\
&=\frac{\displaystyle{c_1\int_0^a S_1(t|{\boldsymbol x})dt+c_j\frac{S_1(a|{\boldsymbol x})}
		{S_j(a|{\boldsymbol x})}\int_a^\eta S_j(t|{\boldsymbol x})dt-c_1(\int_0^a S_1(t|{\boldsymbol x})dt+\frac{S_1(a|{\boldsymbol x})}
		{S_1(a|{\boldsymbol x})}\int_a^\eta S_1(t|{\boldsymbol x})dt)}}
{\displaystyle{\int_0^a S_1(t|{\boldsymbol x})dt+\frac{S_1(a|{\boldsymbol x})}{S_j(a|{\boldsymbol x})}\int_a^\eta S_j(t|{\boldsymbol x})dt-
		(\int_0^a S_1(t|{\boldsymbol x})dt+\frac{S_1(a|{\boldsymbol x})}{S_1(a|{\boldsymbol x})}\int_a^\eta S_1(t|{\boldsymbol x})dt)}} \\
&=\frac{\displaystyle{\frac{c_j}{S_j(a|{\boldsymbol x})}\int_a^\eta S_j(t|{\boldsymbol x})dt-\frac{c_1}{S_1(a|{\boldsymbol x})}
		\int_a^\eta S_1(t|{\boldsymbol x})dt}}
{\displaystyle{\frac{1}{S_j(a|{\boldsymbol x})}\int_a^\eta S_1(t|{\boldsymbol x})dt-\frac{1}{S_1(a|{\boldsymbol x})}
		\int_a^\eta S_1(t|{\boldsymbol x})dt}}= ICER^{STRT}({\boldsymbol x},a). \\
\end{split}
\]
\[
\begin{split}
INB^{DLY}({\boldsymbol x})&=S_1(a|\boldsymbol x)\biggl[\theta\left(\frac{1}{S_2(a|{\boldsymbol x})}\int_a^\eta S_2(t|{\boldsymbol x})dt-
\frac{1}{S_1(a|{\boldsymbol x})}\int_a^\eta S_1(t|{\boldsymbol x})dt\right) \\
&-\left(\frac{c_2}{S_2(a|{\boldsymbol x})}\int_a^\eta S_2(t|{\boldsymbol x})dt-\frac{c_1}
{S_1(a|{\boldsymbol x})}\int_a^\eta S_1(t|{\boldsymbol x})dt\right)\biggl]\\
&= S_1(a|{\boldsymbol x})INB^{STRT}({\boldsymbol x}). \\
\end{split}
\]

\newpage
\section*{Supplementary Appendix 4. Derivation of the theoretical results for RMST and ICER  in the simulation study}
\label{E}

In our simulation setting $\boldsymbol{\chi}$ is a binary covariate and $\lambda_{0j}(t)$ is a constant function, which we will denote here as $\lambda_j$.
Thus, $S_j(t|{X})=\exp\{-\lambda_je^{\beta {X}}t\}$ and
\[
\begin{split}
\bar\mu_{j}^{STRT}(\boldsymbol{\chi},a) &=P(X=1)\int_a^{\eta}e^{-\lambda_j\exp\{\beta\}(t-a)}dt+P(X=0)\int_a^{\eta}e^{-\lambda_j(t-a)}dt\\
&=P(X=1)\frac{e^{-\beta}}{\lambda_j}(1-e^{-\lambda_j\exp{\{\beta\}}(\eta-a)})+P(X=0)\frac{1}{\lambda_j}(1-e^{-\lambda_j(\eta-a)}),\\
\end{split}
\]
\[
\begin{split}
\bar\mu_{j}^{DLY}(\boldsymbol{\chi},a) &=P(X=1)\big[\int_0^a e^{-exp\{\beta\}t}dt+e^{-exp\{\beta\}a}\int_a^{\eta}e^{-\lambda_j\exp\{\beta\}(t-a)}dt\big]\\
&+P(X=0)\big[\int_0^a e^{-t}dt+e^{-a}\int_a^{\eta}e^{-\lambda_j(t-a)}dt\big]=\\
&=P(X=1)\big[ e^{-\beta}(1-e^{-exp\{\beta\}a})+\frac{e^{-\exp\{\beta\}a-\beta}}{\lambda_j}(1-e^{-\lambda_jexp\{\beta\}(\eta-a)})\big] \\
&+P(X=0)\big[1-e^{-a}+\frac{e^{-a}}{\lambda_j}(1-e^{-\lambda_j(\eta-a)})\big],\\
\end{split}
\]

{\color{black}
\[
\begin{split}
\bar\mu_{j}^{DLY}(\boldsymbol{\chi}, \boldsymbol{\Delta}) &= \bar\mu_{j,0}^{DST}(\boldsymbol{\chi}, \boldsymbol{\Delta}) + \bar\mu_{j,1}^{DLY}(\boldsymbol{\chi}, \boldsymbol{\Delta}), \\
\bar\mu_{j,0}^{DLY}(\boldsymbol{\chi}, \boldsymbol{\Delta}) &= P(X=1) \big[ e^{-\beta}(1-e^{-exp\{\beta\}a})\big]/ \lambda_0 + P(X=0)\big(1-e^{-a}\big)/ \lambda_0, \\
\bar\mu_{j,1}^{DLY}(\boldsymbol{\chi}, \boldsymbol{\Delta}) &= P(X=0)\frac{e^{-a}}{\lambda_j} (1-e^{-\lambda_j(\eta-a)})
+ P(X=1) \frac{e^{-\exp\{\beta\}a-\beta}}{\lambda_j} (1-e^{-\lambda_jexp\{\beta\}(\eta-a)}), \\
ICER^{DLY}(\boldsymbol{\chi}, \boldsymbol{\Delta}) &= \frac{c_j\bar\mu_{j,1}^{DLY}(\boldsymbol{\chi}, \boldsymbol{\Delta}) + c_1\mu_{j,0}^{DLY}(\boldsymbol{\chi}, \boldsymbol{\Delta}) - c_1\mu_{1}^{DLY}(\boldsymbol{\chi}, \boldsymbol{\Delta})}{\bar\mu_{j}^{DLY}(\boldsymbol{\chi}, \boldsymbol{\Delta}) - \mu_{1}^{DLY}(\boldsymbol{\chi}, \boldsymbol{\Delta})}, \\
INB^{DLY}(\boldsymbol{\chi}, \boldsymbol{\Delta}) &= \theta(\bar\mu_{j}^{DLY}(\boldsymbol{\chi}, \boldsymbol{\Delta}) - \mu_{1}^{DLY}(\boldsymbol{\chi}, \boldsymbol{\Delta}))
- (c_j\bar\mu_{j,1}^{DLY}(\boldsymbol{\chi}, \boldsymbol{\Delta}) + c_1\mu_{j,0}^{DLY}(\boldsymbol{\chi}, \boldsymbol{\Delta}) \\
&- c_1 \mu_{1}^{DLY}(\boldsymbol{\chi}, \boldsymbol{\Delta}))
\end{split}
\]
}

\[
\bar\mu_{j}^{DST}(\boldsymbol{\chi},\boldsymbol{\Delta})=\sum_{i=1}^{m}Pr(D=\delta_i)\mu_{j}^{DLY}(\boldsymbol{\chi},\delta_i),
\]
where $\Delta$ is a delay distribution, and $D$ is a random variable that represents delay with possible values $\delta_1$, ..., $\delta_m$, for example, as presented in Web Supplementary Figure 3.

As $\eta\rightarrow\infty$, $\widehat{\text{RMST}}$ and $\widehat{\text{ICER}}$ under scenarios STRT and DLY will have the following limits
\[
\bar{\mu}_{j}^{STRT}(\boldsymbol{\chi},a)\xrightarrow[\eta\rightarrow\infty]{}P(X=1)\frac{e^{-\beta}}{\lambda_j}+P(X=0)\frac{1}{\lambda_j},
\]
\[
\bar\mu_{j}^{DLY}(\boldsymbol{\chi},a)\xrightarrow[\eta\rightarrow\infty]{}P(X=1)\big{\{}e^{-\beta}\big[1+e^{-\exp\{\beta\}a}(\frac{1}{\lambda_j}-1)\big]\big{\}}
+P(X=0)\big{\{}1+e^{-a}(\frac{1}{\lambda_j}-1)\big{\}};
\]
with $\lambda_1=1$
\[
ICER^{STRT}\xrightarrow[\eta\rightarrow\infty]{}\frac{c_2-c_1\lambda_2}{1-\lambda_2},
\]

\small
\begin{align*}
ICER^{DLY}&\xrightarrow[\eta\rightarrow\infty]{}
\frac{\displaystyle{c_2\left(P(X=1)\frac{e^{-e^{\beta}a-\beta}}{\lambda_2}+P(X=0)\frac{e^{-a}}{\lambda_2}\right)-
		c_1\left(P(X=1)e^{-e^{\beta}a-\beta}+P(X=0)e^{-a}\right)}}
{\displaystyle{P(X=1)\frac{e^{-e^{\beta}a-\beta}}{\lambda_2}+P(X=0)\frac{e^{-a}}{\lambda_2}-
		\left(P(X=1)e^{-e^{\beta}a-\beta}+P(X=0)e^{-a}\right)}}
\\&=\frac{c_2-c_1\lambda_2}{1-\lambda_2},
\end{align*}
\normalsize
and
\[
ICER^{DST}\xrightarrow[\eta\rightarrow\infty]{}\frac{\displaystyle{\sum_{i=1}^{m}P(D=\delta_i)\left(P(X=1)e^{-e^{\beta}a-\beta}+
		P(X=0)e^{-a}\right)\left(\frac{c_2}{\lambda_2}-c_1\right)}}
{\displaystyle{\sum_{i=1}^{m}P(D=\delta_i)\left(P(X=1)e^{-e^{\beta}a-\beta}+P(X=0)e^{-a}\right)\left(\frac{1}{\lambda_2}-1\right)}}=\frac{c_2-c_1\lambda_2}{1-\lambda_2}.
\]

\newpage
\section*{Supplementary Appendix 5. Additional simulation results}
\label{F}

We hereby report additional simulation results for settings with different sample sizes and baseline hazards.

\begin{table}[h]
\caption{Simulation study results: Performance of $\widehat{\text{RMST}}$ under scenarios STRT and DLY.}
\centering
\footnotesize {
\begin{tabular}{cccccccc}
\hline
& & \multicolumn{2}{c}{\% Relative Bias} & Coverage & \multicolumn{2}{c}{\% Relative Bias} & Coverage \\
Scenario & Delay (\%) & $\widehat{\mu}_1$ & $\widehat{\text{SE}}$(${\mu}_1$) & Probability & $\widehat{\mu}_2$	& $\widehat{\text{SE}}$(${\mu}_2$) & Probability \\
\hline
\noalign{\medskip}
No delay & 0 & 0.4 & 6.8 & 0.94 & 0.3 & 3.8 & 0.94 \\
\noalign{\medskip}
STRT & 10 & -0.1 & 4.8 & 0.95 & -0.3 & 4.3 & 0.93 \\
& 50 & 0.2 & 4.4 & 0.95 & -0.2 & 3.2 & 0.94 \\
\noalign{\medskip}
DLY & 10 & 0.1 & 13.3 & 0.94 & -0.1 &	11.1 & 0.95 \\
& 50 & 0.4 & 12.8 & 0.94 & 0.0 & 9.7 & 0.96 \\
\hline
\multicolumn{8}{l}{Sample size $500$, baseline hazard 1, hazard ratio 0.5, and number of simulation replicates 1000.}  \\
\multicolumn{8}{l}{Controlling for one Bernoulli covariate generated with probability for success 0.9.}
\end{tabular}
}
\end{table}

\begin{table}[h]
\caption{Simulation study results: Performance of $\widehat{\text{ICER}}$ and $\widehat{\text{INB}}^*$ under scenarios STRT and DLY.}
\centering
\footnotesize {
\begin{tabular}{cccccccc}
\hline
& & \multicolumn{2}{c}{\% Relative Bias} & Coverage & \multicolumn{2}{c}{\% Relative Bias} & Coverage \\
Scenario & Delay (\%) & $\widehat{\text{ICER}}$ & $\widehat{\text{SE}}$(${\text{ICER}}$) & Probability & $\widehat{\text{INB}}$	& $\widehat{\text{SE}}$(${\text{INB}}$) & Probability \\
\hline
\noalign{\medskip}
No delay & 0 & 2.5 & -5.5 & 0.95 & -0.3 & 0.4 & 0.95 \\
\noalign{\medskip}
STRT & 10 & 3.5 & -6.7 & 0.95 & -2.4 & 0.1 & 0.95 \\
& 50 & 4.1 & -6.8 & 0.95 & -4.2 & -0.2 & 0.95 \\
\noalign{\medskip}
DLY & 10 & 2.3 & -5.8 & 0.94 & 3.1 & 1.1 & 0.95 \\
& 50 & 1.8 & -5.2 & 0.93 & 5.3 & 2.0 & 0.95 \\
\hline
\multicolumn{8}{l}{Sample size $500$, baseline hazard 1, hazard ratio 0.5, and number of simulation replicates 1000.}  \\
\multicolumn{8}{l}{Controlling for one Bernoulli covariate generated with probability for success 0.9.} 
\end{tabular}
}
\end{table}

\begin{landscape}
\begin{table}[]
\small
\caption{Simulation study results: Performance of $\widehat{\text{RMST}}$ under scenarios STRT and DLY.}
\centering
{\footnotesize 
\begin{tabular}{cccccccccc}
\hline
Sample & Baseline & & & \multicolumn{2}{c}{\% Relative Bias} & Coverage & \multicolumn{2}{c}{\% Relative Bias} & Coverage \\
Size & Hazard & Scenario & Delay (\%) & $\widehat{\mu}_1$ & $\widehat{\text{SE}}$(${\mu}_1$) & Probability & $\widehat{\mu}_2$	& $\widehat{\text{SE}}$(${\mu}_2$) & Probability \\
\hline
\noalign{\medskip}
500  & 0.8 & No delay & 0     & 0.1\%     & 4.5\%       & 0.94       & 0.0\%     & 4.1\%       & 0.94       \\
& & STRT     & 10    & -0.3\%    & 3.2\%       & 0.94       & -0.7\%    & 3.8\%       & 0.93       \\
& & & 50    & -0.1\%    & 3.2\%       & 0.94       & -0.6\%    & 3.3\%       & 0.94       \\
& & DLY      & 10    & -0.3\%    & 10.5\%      & 0.95       & -0.6\%    & 11.9\%      & 0.95       \\
& & & 50    & 0.0\%     & 10.5\%      & 0.94  & -0.5\%    & 10.1\%      & 0.95 \\
& 1.2 & No delay & 0     & 0.0\%     & 4.4\%       & 0.95       & 0.0\%     & 4.0\%       & 0.93       \\
& & STRT     & 10    & -0.4\%    & 4.6\%       & 0.94       & -0.5\%    & 5.5\%       & 0.93       \\
& & & 50    & 0.0\%     & 4.6\%       & 0.94       & -0.3\%    & 4.7\%       & 0.94       \\
& & DLY      & 10    & -0.2\%    & 10.7\%      & 0.95       & -0.4\%    & 8.7\%       & 0.96       \\
& & & 50    & 0.1\%     & 10.5\%      & 0.95       & -0.2\%    & 7.1\%       & 0.96\\  
\noalign{\medskip}
1000 & 0.8 & no delay & 0     & 0.1\%      & 3.6\%       & 0.93      & 0.1\%      & 5.3\%  & 0.94 \\
& & STRT     & 10    & -0.1\%     & 2.1\%       & 0.94      & -0.2\%     & 6.4\%  & 0.93 \\
& & & 50    & 0.1\%      & 2.3\%       & 0.94      & -0.1\%     & 7.2\%  & 0.93 \\
& & DLY      & 10    & 0.0\%      & 9.3\%       & 0.93      & -0.1\%     & 14.9\% & 0.96 \\
& & & 50    & 0.2\%      & 9.1\%       & 0.93      & 0.0\%      & 15.2\% & 0.96 \\
& 1.2 & no delay & 0     & -0.1\%     & 6.2\%       & 0.94      & 0.1\%      & 2.1\%  & 0.95 \\
& & STRT     & 10    & -0.2\%     & 5.0\%       & 0.93      & -0.1\%     & 1.4\%  & 0.95 \\
& & & 50    & 0.1\%      & 4.6\%       & 0.93      & 0.1\%      & 1.5\%  & 0.94 \\
& & DLY      & 10    & -0.1\%     & 12.8\%      & 0.93      & 0.0\%      & 10.1\% & 0.96 \\
& & & 50    & 0.2\%      & 12.4\%      & 0.93      & 0.3\%      & 9.8\%  & 0.95\\
\noalign{\medskip}
10000 & 0.8 & No delay & 0     & 0.0\%     & 2.2\%       & 0.94       & 0.0\%     & 2.3\%       & 0.92       \\
& & STRT     & 10    & 0.1\%     & 2.9\%       & 0.95       & 0.1\%     & 3.2\%       & 0.92       \\
& & & 50    & 0.3\%     & 2.7\%       & 0.93       & 0.3\%     & 3.0\%       & 0.91       \\
& & DLY      & 10    & 0.1\%     & 8.9\%       & 0.94       & 0.1\%     & 9.6\%       & 0.95       \\
& & & 50    & 0.3\%     & 8.1\%       & 0.93       & 0.4\%     & 8.6\%       & 0.93       \\
& 1.2 & No delay & 0     & 0.0\%     & 6.1\%       & 0.94       & 0.0\%     & 6.0\%       & 0.93       \\
& & STRT     & 10    & 0.1\%     & -1.1\%      & 0.96       & 0.1\%     & -1.1\%      & 0.92       \\
& & & 50    & 0.4\%     & 3.7\%       & 0.93       & 0.5\%     & 3.7\%       & 0.90       \\
& & DLY      & 10    & 0.1\%     & 12.1\%      & 0.94       & 0.1\%     & 11.6\%      & 0.95       \\
& & & 50    & 0.4\%     & 8.9\%       & 0.93  & 0.5\%     & 8.5\%       & 0.93  \\
\hline
\multicolumn{10}{l}{Scenario STRT: for patients who survive to a fixed time $r=0.5$; Scenario DLY: to a fixed delay time $a=0.5$;} \\
\multicolumn{10}{l}{Hazard ratio 0.5 and number of simulation replicates 1000, controlling for one Bernoulli covariate with probability 0.9.}
\end{tabular}
}
\end{table}
\end{landscape}

\begin{landscape}
\begin{table}[]
\small
\caption{Simulation results: Performance of $\widehat{\text{ICER}}$ and $\widehat{\text{INB}}^*$ under scenarios STRT and DLY.}
{\footnotesize
\begin{tabular}{llllllll}
\hline
Sample & Baseline & & & \% Relative Bias & Coverage & \% Relative Bias & Coverage \\
Size & Hazard & Scenario & Delay (\%) & $\widehat{\text{ICER}}$ & Probability & $\widehat{\text{INB}}$ & Probability \\
\hline
\noalign{\medskip}
500 & 0.8 & No delay & 0     & 2.8\%      & 0.94        & -0.5\%    & 0.96       \\
& & STRT     & 10    & 4.6\%      & 0.93        & -6.9\%    & 0.96       \\
& & & 50    & 5.3\%      & 0.95        & -10.3\%   & 0.96       \\
& & DLY      & 10    & 2.6\%      & 0.93        & 4.1\%     & 0.96       \\
& & & 50    & 2.0\%      & 0.92        & 8.6\%     & 0.95       \\
& 1.2 & No delay & 0     & 2.1\%      & 0.95        & -0.1\%    & 0.95       \\
& & STRT     & 10    & 2.9\%      & 0.94        & -1.6\%    & 0.94       \\
& & & 50    & 3.3\%      & 0.94        & -2.5\%    & 0.95       \\
& & DLY      & 10    & 2.1\%      & 0.93        & 1.7\%     & 0.95       \\
& & & 50    & 1.7\%      & 0.93        & 3.4\%     & 0.95   \\  
\noalign{\medskip}
1000 & 0.8 & No delay & 0     & 1.4\%      & 0.95        & 0.2\%     & 0.95       \\
& & STRT     & 10    & 2.0\%      & 0.96        & -1.8\%    & 0.96       \\
& & & 50    & 2.5\%      & 0.96        & -4.1\%    & 0.95       \\
& & DLY      & 10    & 1.0\%      & 0.95        & 3.9\%     & 0.96       \\
& & & 50    & 0.8\%      & 0.95        & 6.0\%     & 0.96       \\
& 1.2 & No delay & 0     & 0.6\%      & 0.94        & 1.4\%     & 0.94       \\
& & STRT & 10    & 0.8\%      & 0.94        & 1.1\%     & 0.95       \\
& & & 50    & 1.1\%      & 0.94        & 0.5\%     & 0.95       \\
& & DLY & 10    & 0.5\%      & 0.94        & 2.7\%     & 0.95       \\
& & & 50    & 0.4\%      & 0.93        & 3.5\%     & 0.95       \\
\noalign{\medskip}
10000 & 0.8 & No delay & 0 & 0.1\% & 0.95    & 0.1\% & 0.95 \\
& & STRT & 10 & 0.1\% & 0.95 & 0.1\% & 0.95 \\
& & & 50 & 0.3\% & 0.95 & -0.5\% & 0.94 \\
& & DLY & 10 & 0.1\% & 0.95 & 0.5\% & 0.95 \\
& & & 50 & 0.1\% & 0.95 & 0.7\% & 0.95 \\
& 1.2 & No delay & 0 & 0.1\% & 0.93 & 0.0\%  & 0.93 \\
& & STRT & 10 & 0.2\% & 0.95 & -0.1\% & 0.94 \\
& & & 50 & 0.3\% & 0.94 & -0.1\% & 0.93 \\
& & DLY & 10 & 0.2\% & 0.94 & 0.0\% & 0.94 \\
& & & 50 & 0.1\% & 0.95 & 0.4\% & 0.94  \\
\hline
\multicolumn{8}{l}{Scenario STRT: Patients who survive to a fixed time $r=0.5$; Scenario DLY: to a fixed delay time $a=0.5$;}\\
\multicolumn{8}{l}{Hazard ratio 0.5 and number of simulation replicates 1000, controlling for one Bernoulli covariate with probability 0.9.}
\end{tabular}}
\end{table}
\end{landscape}

\begin{landscape}
\begin{table}[h]
\small
\caption{Simulation results: Performance of $\widehat{\text{RMST}}$, $\widehat{\text{ICER}}$, and $\widehat{\text{INB}}^*$ under scenarios STRT and DLY.}
{\footnotesize
\begin{tabular}{lllllllllll}
\hline
Baseline & & & $\widehat{\mu}_1$ & $\widehat{\mu}_1$ & $\widehat{\mu}_2$ & $\widehat{\mu}_2$ & $\widehat{\text{ICER}}$ & $\widehat{\text{ICER}}$ & $\widehat{\text{INB}}^*$ & $\widehat{\text{INB}}^*$ \\
Hazard & Scenario & Delay (\%) & \% RBias & Empirical SE & \% RBias & Empirical SE & \% RBias & Empirical SE & \% RBias & Empirical SE \\
\hline
\noalign{\medskip}
5 & No delay & 0 & 0.9\%  & 0.18  & -0.2\%  & 0.28 & 1.1\% & 36.6  & -1.0\% & 172  \\
& STRT & 10  & 1.1\% & 0.18 & 1.0\% & 0.28        & 0.6\% & 36.6 & 0.9\% & 173 \\
&  & 50  & 1.9\% & 0.19 & 1.8\% & 0.31        & 0.6\% & 36.8 & 1.7\% & 180 \\
& DLY & 10  & 0.3\% & 0.06 & 0.5\% & 0.10        & 0.5\% & 35.5 & 0.9\% & 112 \\
&  & 50  & 1.1\% & 0.06 & 1.4\% & 0.11        & 0.4\% & 35.5 & 2.0\% & 114 \\
\noalign{\medskip}
10 & No delay & 0 & 1.1\%  & 0.09  & 1.1\%  & 0.17 & 0.8\% & 39.2  & 1.2\% & 117  \\
& STRT & 10  & 1.2\% & 0.09 & 1.4\% & 0.18 & 0.7\%    & 39.2 & 1.6\% & 118 \\
&  & 50  & 2.2\% & 0.10 & 2.3\%  & 0.19 & 0.8\%   & 39.9 & 2.6\% & 126 \\
& DLY & 10  & 0.3\% & 0.03 & 0.4\% & 0.05 & 0.7\%    & 38.9 & 0.8\% & 50.1 \\
&  & 50  & 1.3\% & 0.03 & 1.5\%  & 0.05 & 0.8\%   & 39.6 & 2.0\% & 51.7 \\
\noalign{\medskip}
20 & No delay & 0 & 0.9\%  & 0.05  & 1.1\%  & 0.09 & 1.3\% & 53.8  & 1.4\% & 70.4 \\
& STRT & 10  & 1.0\% & 0.52 & 1.1\% & 0.09 & 1.3\%    & 54.2 & 1.4\% & 71.2  \\
&  & 50  & 1.8\% & 0.05 & 2.1\% & 0.10 & 1.3\%   & 55.3 & 2.6\% & 76.8 \\
& DLY & 10  & 0.2\% & 0.02 & 0.4\% & 0.02 & 1.3\%    & 54.1 & 0.8\% & 16.6  \\
&  & 50  & 1.2\% & 0.02 & 1.3\% & 0.02 & 1.3\%   & 54.9 & 2.0\% & 17.0 \\
\hline
\multicolumn{10}{l}{Scenario STRT: Patients who survive to a fixed time $r=0.5$; Scenario DLY: to a fixed delay time $a=0.5$;}\\
\multicolumn{10}{l}{Sample size $1000$, hazard ratio 0.5, and number of simulation replicates 1000.} \\
\multicolumn{10}{l}{Controlling for one Bernoulli covariate generated with probability of success 0.9.}
\end{tabular}}
\end{table}
\end{landscape}

\newcommand{\multilinecell}[2][c]{
\begin{tabular}[#1]{@{}c@{}}#2\end{tabular}}
\begin{table}
\centering
\caption{Censoring rate and proportion of patients not receiving new treatment under different simulation settings.}
{\scriptsize
\begin{tabular}{cccccc}
\hline
\multilinecell{Baseline \\ Hazard} & Delay (\%) & HR & \multilinecell{Censoring Rate \\($j=1$)} & \multilinecell{Censoring Rate\\($j=2$)} & \multilinecell{\% Missing New Treatment\\($j=2$)} \\ 
\hline
0.8 & 0 & 0.2 & 0.33 & 0.76 & 0.00 \\ 
& & 0.5 & 0.33 & 0.55 & 0.00 \\ 
& & 0.8 & 0.33 & 0.40 & 0.00 \\ 
& 10 & 0.2 & 0.33 & 0.76 & 0.00 \\ 
& & 0.5 & 0.33 & 0.54 & 0.01 \\ 
& & 0.8 & 0.33 & 0.40 & 0.01 \\ 
& 50 & 0.2 & 0.33 & 0.75 & 0.02 \\ 
& & 0.5 & 0.33 & 0.54 & 0.04 \\ 
& & 0.8 & 0.33 & 0.40 & 0.07 \\ 
1.0 & 0 & 0.2 & 0.26 & 0.71 & 0.00 \\ 
& & 0.5 & 0.26 & 0.48 & 0.00 \\ 
& & 0.8 & 0.26 & 0.33 & 0.00 \\ 
& 10 & 0.2 & 0.26 & 0.71 & 0.01 \\ 
& & 0.5 & 0.26 & 0.48 & 0.01 \\ 
& & 0.8 & 0.26 & 0.33 & 0.02 \\ 
& 50 & 0.2 & 0.26 & 0.71 & 0.03 \\ 
& & 0.5 & 0.26 & 0.48 & 0.05 \\ 
& & 0.8 & 0.26 & 0.33 & 0.08 \\ 
1.2 & 0 & 0.2 & 0.20 & 0.67 & 0.00 \\ 
& & 0.5 & 0.20 & 0.42 & 0.00 \\ 
& & 0.8 & 0.20 & 0.27 & 0.00 \\ 
& 10 & 0.2 & 0.20 & 0.67 & 0.01 \\ 
& & 0.5 & 0.20 & 0.42 & 0.01 \\ 
& & 0.8 & 0.20 & 0.27 & 0.02 \\ 
& 50 & 0.2 & 0.20 & 0.67 & 0.03 \\ 
& & 0.5 & 0.20 & 0.42 & 0.06 \\ 
& & 0.8 & 0.20 & 0.27 & 0.10 \\ 
5 & 0 & 0.2 & 0.01 & 0.26 & 0.00 \\ 
& & 0.5 & 0.01 & 0.05 & 0.00 \\ 
& & 0.8 & 0.01 & 0.02 & 0.00 \\ 
& 10 & 0.2 & 0.01 & 0.26 & 0.02 \\ 
& & 0.5 & 0.01 & 0.05 & 0.04 \\ 
& & 0.8 & 0.01 & 0.02 & 0.06 \\ 
& 50 & 0.2 & 0.01 & 0.26 & 0.10 \\ 
& & 0.5 & 0.01 & 0.05 & 0.20 \\ 
& & 0.8 & 0.01 & 0.02 & 0.28 \\ 
10 & 0 & 0.2 & 0.01 & 0.08 & 0.00 \\ 
& & 0.5 & 0.01 & 0.01 & 0.00 \\ 
& & 0.8 & 0.01 & 0.01 & 0.00 \\ 
& 10 & 0.2 & 0.01 & 0.08 & 0.03 \\ 
& & 0.5 & 0.01 & 0.01 & 0.07 \\ 
& & 0.8 & 0.01 & 0.01 & 0.09 \\ 
& 50 & 0.2 & 0.01 & 0.08 & 0.17 \\ 
& & 0.5 & 0.01 & 0.01 & 0.32 \\ 
& & 0.8 & 0.01 & 0.01 & 0.44 \\ 
20 & 0 & 0.2 & 0.00 & 0.02 & 0.00 \\ 
& & 0.5 & 0.00 & 0.01 & 0.00 \\ 
& & 0.8 & 0.00 & 0.00 & 0.00 \\ 
& 10 & 0.2 & 0.00 & 0.02 & 0.06 \\ 
& & 0.5 & 0.00 & 0.01 & 0.10 \\ 
& & 0.8 & 0.00 & 0.00 & 0.12 \\ 
& 50 & 0.2 & 0.00 & 0.02 & 0.28 \\ 
& & 0.5 & 0.00 & 0.01 & 0.50 \\ 
& & 0.8 & 0.00 & 0.00 & 0.62 \\ 
\hline
\multicolumn{6}{l}{$j$ is the index for the treatment.} \\
\multicolumn{6}{l}{\% Missing New Treatment: \% of delayed patients missing Treatment 2 due to death during the delay.}
\\
\multicolumn{6}{l}{Our motivating data is closest to the setting with low mortality, HR 0.5, and 50\% delay.}
\end{tabular}}
\end{table}
\clearpage

\pagebreak
\newpage
\section*{\textcolor{black}{Supplementary Appendix 6. Further graphics and data summary of HIV clinics}}
\label{S}

\begin{figure*}[h]{Suppenmentary Figure 1. Options for delay time distribution considered in the simulations for Scenario DST.
Option I uniform, option II right skewed, and option III left skewed.}
\begin{center}
\includegraphics[height=3.4 in]{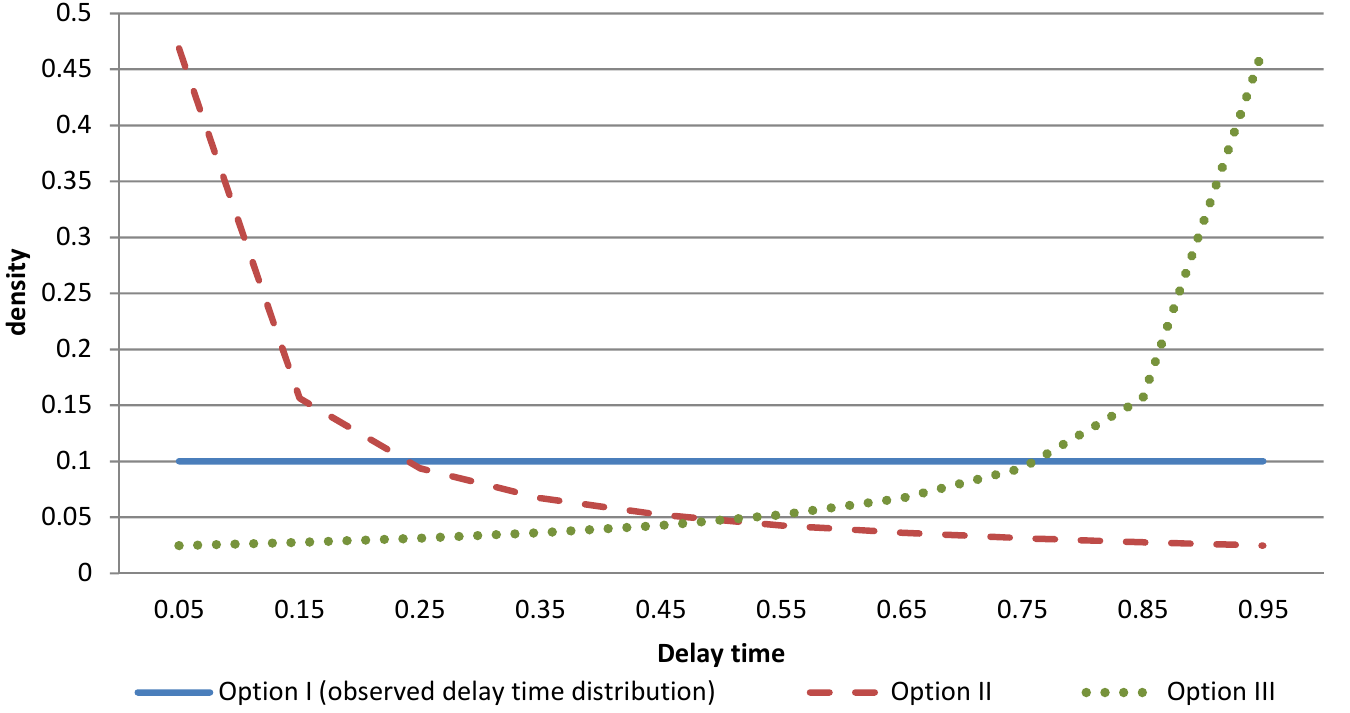}
\end{center}
\end{figure*}

\begin{figure*}{Suppenmentary Figure 2. ICER as a functions of the follow-up periods, $\eta$ calculated under STRT scenario }
\begin{center}
\includegraphics[height=2.8 in]{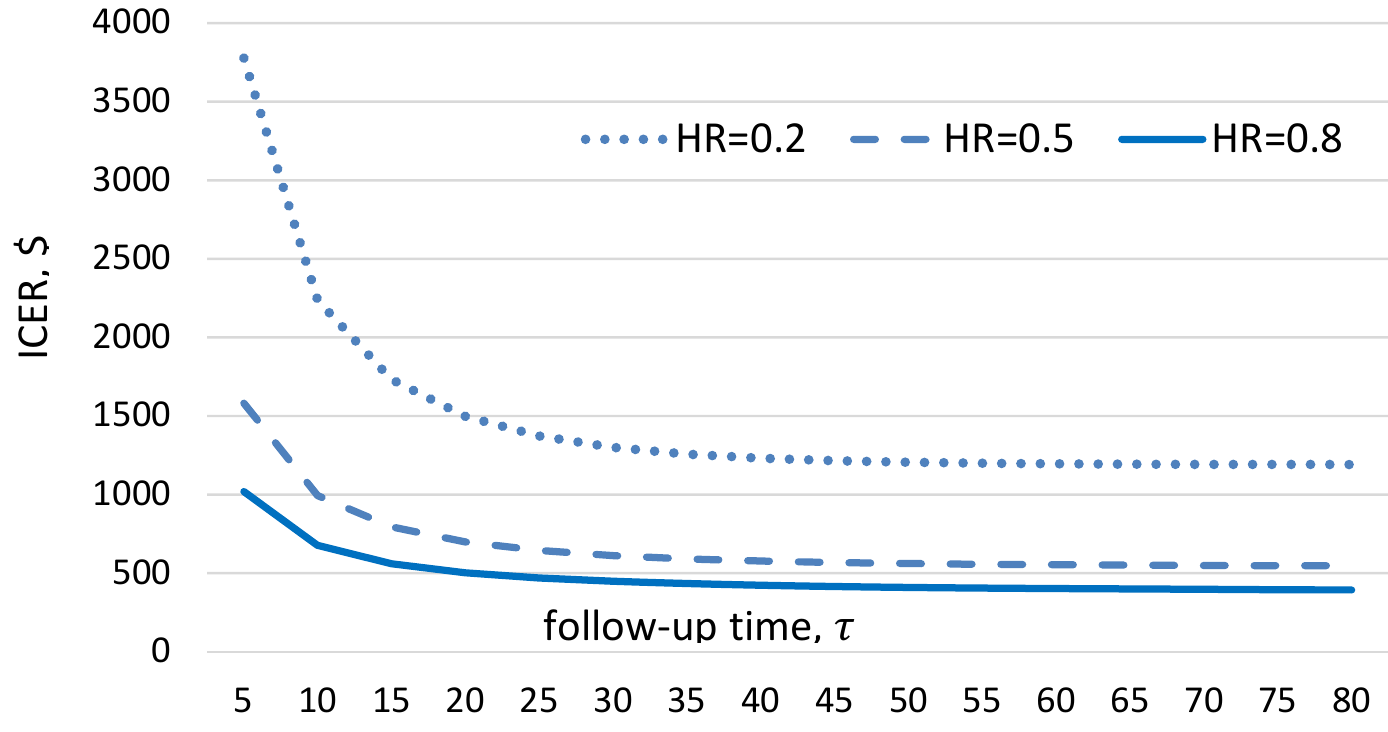}
\end{center}
\end{figure*}

\begin{figure*}{Suppenmentary Figure 3. Data flow of ARV2 eligible patients, Dar es Salaam, Tnazania, 2004-2012.}
\begin{center}
\includegraphics[height=2.5in]{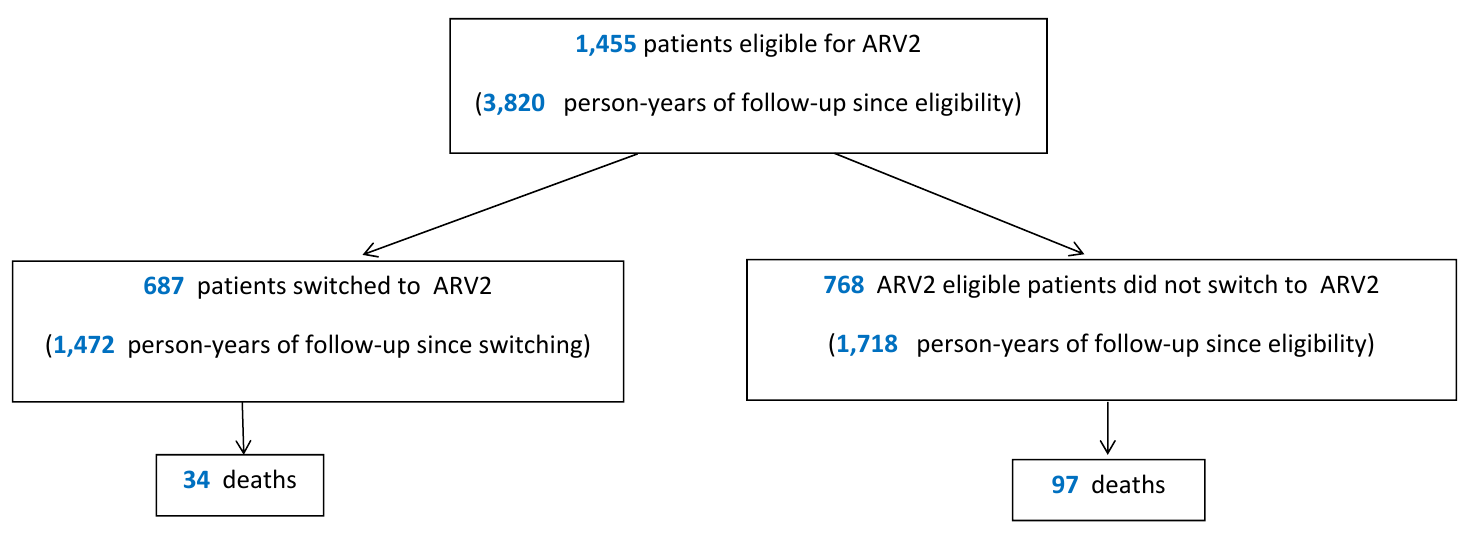}
\end{center}
\end{figure*}

\begin{figure*}{Suppenmentary Figure 4. Delay time distribution$^{1}$ observed among those who switched to ARV2, Dar es Salaam, Tanzania, 2004-2012.}
\begin{center}
\includegraphics[height=3 in]{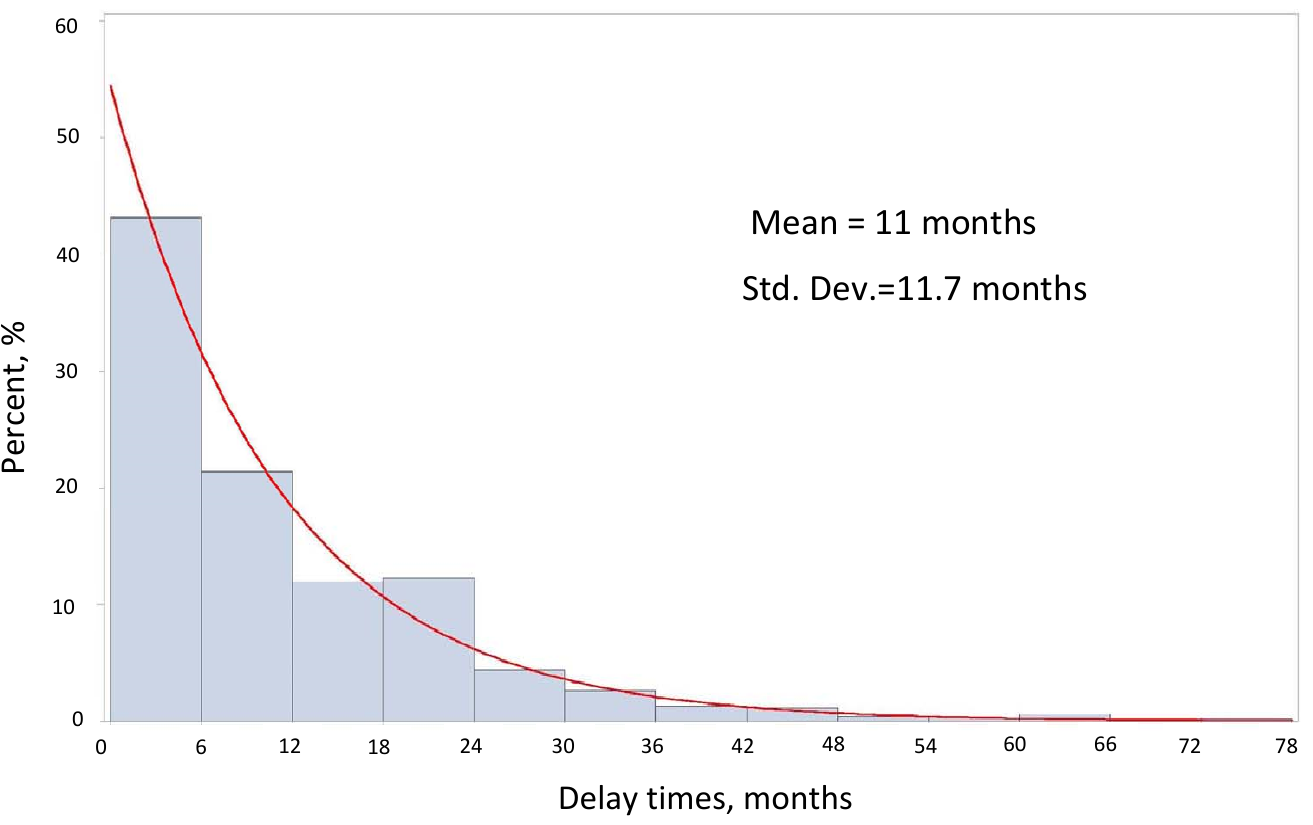}
\end{center}
$^1$ Red curve is an approximated exponential fit for delay time distribution.
\end{figure*}

\begin{table*}
{Supplementary Table 1. Baseline characteristics at ARV2 eligibility among 1,455 ARV2 eligible HIV patients, Dar es Salaam, Tanzania, 2004-2012.}
\begin{center}
{\small 
\begin{tabular}{lc}
& $n$ (\%)             \\
\hline
Male & 497 (34\%)\\
Age categories (years) & \\
$\qquad$ $< 30$ & 258 (18\%)\\
$\qquad$ $30$ to $<40$ & 681 (47\%)\\
$\qquad$ $40$ to $< 50$ & 368 (25\%) \\
$\qquad$ $50+$ & 145 (10\%) \\
Married & 587 (40\%)\\
BMI categories (kg/($\text{m}^2$) & \\
$\qquad$ $< 17$ & 169 (12\%)\\
$\qquad$ $17.0$ to $<18.5$ &  758 (52\%)\\
$\qquad$ $18.5$ to $<25.0$ & 261 (18\%)\\
$\qquad$ $ 25.0+$ & 125 (9\%)\\
Haemoglobin (g/dL) &  \\
$\qquad$ $<$7.5 & 33 (2\%)\\
$\qquad$ 7.5 to $<$10 & 228 (16\%)\\
$\qquad$ 10+ & 1,194 (82\%)\\
CD4+ T cell count categories (cells/$\mu$L)  & \\
$\qquad$ $<$50  & 180 (12\%) \\
$\qquad$ 50 to $<$100  & 225 (16\%) \\
$\qquad$ 100 to $<$200 & 464 (32\%)\\
$\qquad$ 200+    &  586 (40\%) \\
Use of cotrimoxazole &  766 (53\%)\\
HIV stage & \\
$\qquad$ I & 55 (4\%) \\
$\qquad$ II & 156 (11\%) \\
$\qquad$ III & 885 (61\%) \\
$\qquad$ IV & 359 (25\%) \\
History of tuberculosis & 439 (30\%) \\
Facility level & \\
$\qquad$ Hospital & 1274 (88\%) \\
$\qquad$ Health center & 76 (5\%) \\
$\qquad$ Dispensary & 105 (7\%) \\
Season of visit & \\
$\qquad$ Long Dry (June - September) & 592 (41\%) \\
$\qquad$ Short Rainy (October - November) & 194 (13\%) \\
$\qquad$ Short Dry (December - March) & 370 (25\%) \\
$\qquad$ Long Rainy (April - March) & 299 (21\%) \\
\hline
\end{tabular}}
\end{center}
\end{table*}